\documentclass[lettersize,journal]{IEEEtran}
\usepackage{amsmath,amsfonts}
\usepackage{amsmath,amssymb,bm}
\usepackage{mathrsfs}
\usepackage{algorithmic}
\usepackage{algorithm}
\usepackage{array}
\usepackage[caption=false,font=normalsize,labelfont=sf,textfont=sf]{subfig}
\usepackage{textcomp}
\usepackage{stfloats}
\usepackage{url}
\usepackage{verbatim}
\usepackage{graphicx}
\usepackage{cite}
\usepackage{multirow} 
\usepackage{booktabs}
\begin{document}

\title{PCE-GAN: A Generative Adversarial Network for Point Cloud Attribute Quality Enhancement based on Optimal Transport}

\author{Tian Guo, Hui Yuan,~\IEEEmembership{Senior Member,~IEEE,} Qi Liu, Honglei Su,~\IEEEmembership{Member,~IEEE}, Raouf Hamzaoui,~\IEEEmembership{Senior Member,~IEEE}, and Sam Kwong~\IEEEmembership{Fellow,~IEEE}
\thanks{This work was supported in part by the National Natural Science Foundation of China under Grants 62222110, 62172259, and 62311530104, the High-end Foreign Experts Recruitment Plan of Chinese Ministry of Science and Technology under Grant G2023150003L, the Taishan Scholar Project of Shandong Province (tsqn202103001), the Natural Science Foundation of Shandong Province under Grant ZR2022ZD38, and the OPPO Research Fund. \textit{(Corresponding author: Hui Yuan)}} 
\thanks{Tian Guo and Hui Yuan are with the School of Control Science and Engineering, Shandong University, Jinan 250061, China, and also with the Key Laboratory of Machine Intelligence and System Control, Ministry of Education, Ji'nan, 250061, China (e-mail: guotiansdu@mail.sdu.edu.cn; huiyuan@sdu.edu.cn).}
\thanks{Qi Liu and Honglei Su are with the School of Electronic Information, Qingdao University, Qingdao, 266071, China (e-mail: sdqi.liu@gmail.com; suhonglei@qdu.edu.cn).}
\thanks{Raouf Hamzaoui is with the School of Engineering and Sustainable Development, De Montfort University, LE1 9BH Leicester, UK. (e-mail: rhamzaoui@dmu.ac.uk).}
\thanks{Sam Kwong is with the Department of Computing and Decision Science,
 Lingnan University, Hong Kong (e-mail: samkwong@ln.edu.hk).}
}

\markboth{Journal of \LaTeX\ Class Files,~Vol.~14, No.~8, August~2021}%
{Guo \MakeLowercase{\textit{et al.}}: PCE-GAN: A Generative Adversarial Network for Point Cloud Attribute Quality Enhancement based on Optimal Transport}


\maketitle

\begin{abstract}
Point cloud compression significantly reduces data volume but sacrifices reconstruction quality, highlighting the need for advanced quality enhancement techniques. Most existing approaches focus primarily on point-to-point fidelity, often neglecting the importance of perceptual quality as interpreted by the human visual system. To address this issue, we propose a generative adversarial network for point cloud quality enhancement (PCE-GAN), grounded in optimal transport theory, with the goal of simultaneously optimizing both data fidelity and perceptual quality. The generator consists of a local feature extraction (LFE) unit, a global spatial correlation (GSC) unit and a feature squeeze unit. The LFE unit uses dynamic graph construction and a graph attention mechanism to efficiently extract local features, placing greater emphasis on points with severe distortion. The GSC unit uses the geometry information of neighboring patches to construct an extended local neighborhood and introduces a transformer-style structure to capture long-range global correlations. The discriminator computes the deviation between the probability distributions of the enhanced point cloud and the original point cloud, guiding the generator to achieve high quality reconstruction. Experimental results show that the proposed method achieves state-of-the-art performance. Specifically, when applying PCE-GAN to the latest geometry-based point cloud compression (G-PCC) test model, it achieves an average BD-rate of -19.2\% compared with the PredLift coding configuration and -18.3\% compared with the RAHT coding configuration. Subjective comparisons show a significant improvement in texture clarity and color transitions, revealing finer details and more natural color gradients.
\end{abstract}


\section{Introduction}
\IEEEPARstart{P}{oint} clouds are collections of 3D data that represent the surface of objects or scenes, containing geometry coordinates and attribute information such as colors, reflectance and normals \cite{ref1,ref2,ref3}. They are widely used in areas like virtual reality, digital twins, autonomous driving, and cultural heritage protection \cite{ref4,ref5,ref6}, due to their highly flexible representation. However, point clouds present challenges due to their large data volume, disordered structure, and unstructured point distribution, making them difficult to process, store and transmit. Therefore, it is crucial to develop effective point cloud compression techniques.

To promote the standardization of point cloud compression, the 3D graphics group of the Moving Picture Experts Group (MPEG) initiated a call for proposals in 2017 \cite{ref7}. This led to the development of two point cloud compression technologies: geometry-based point cloud compression (G-PCC) \cite{ref8} and video-based point cloud compression (V-PCC)\cite{ref9}. G-PCC processes 3D geometry information and its attributes directly in 3D space. It first compresses the geometry information of the point cloud using techniques such as octree \cite{refa1}, predictive geometry \cite{refa2}, or trisoup \cite{refa3}. Then it compresses the attribute information based on the reconstructed geometry using region-adaptive hierarchical transform (RAHT) \cite{ref10} or predictive/lifting Transform (PredLift) \cite{ref11}. In contrast, V-PCC transforms 3D geometry and its attributes into 2D video representations and uses advanced video compression standards such as H.265/HEVC \cite{ref12} or H.266/VVC \cite{refa4} to compress them. Although G-PCC and V-PCC achieve remarkable performance, lossy compression inevitably introduces significant attribute distortion, which not only drastically reduces data fidelity but also results in poor subjective visual experience. Therefore, in this paper, we study the attribute quality enhancement method for compressed point clouds.

In recent years, there has been a steady stream of research aimed at enhancing the quality of point cloud attributes. Traditional methods \cite{refa5,ref13,ref14,ref15,refa6,refa7,ref16,ref17,ref18} can be broadly divided into two categories: classical filter-based techniques and graph signal processing-based techniques. However, the former has limited performance in dealing with non-linear distortions, while the latter relies heavily on the way the graph is constructed. Recently, deep learning-based techniques \cite{ref19,ref20,ref21,refd0,ref22,refd1,ref23,ref24,ref25} were proposed to significantly improve the quality of reconstructed point clouds through graph convolution-based methods, sparse convolution-based methods, and projection-based methods.

However, the aforementioned methods address only point-to-point distortion and neglect the perception quality of the human visual system, resulting in suboptimal performance in delivering high-quality visual reconstruction. Therefore, we propose a point cloud attribute quality enhancement algorithm based on optimal transport theory, which achieves high perceptual quality by constraining the enhanced attributes to have the same distribution as the original ones, while simultaneously guaranteeing data fidelity through minimum-distance transport to achieve high objective quality. Specifically, we model the quality enhancement problem as an optimal transport problem and propose a generative adversarial network (GAN) to solve it. In the proposed GAN, the generator mainly consists of three units: a local feature extraction (LFE) unit, a global spatial correlation (GSC) unit, and a feature squeeze (FS) unit. Meanwhile, the discriminator uses a Wasserstein generative adversarial network with gradient penalty (WGAN-GP) \cite{ref26} to compute the Wasserstein distance between the enhanced point cloud and the original point cloud. The aim is to constrain the distribution differences between the original and enhanced point clouds. In detail, the contributions of this paper are summarized as follows.
\begin{itemize}
\item{We propose a point cloud attribute quality enhancement algorithm that considers both perceptual quality and data fidelity by constraining the enhanced point cloud to have the same distribution as the original point cloud and maximizing the mutual information between the enhanced point cloud and the original point cloud. To the best of our knowledge, this is the first exploration of point cloud attribute quality enhancement that considers perceptual quality explicitly.}
\item{We model the quality enhancement problem as an optimal transport problem and develop a relaxation strategy to solve it. Then, we propose a quality enhancement network to solve this problem. The network introduces a WGAN-GP structure to compute the Wasserstein distance between the enhanced and original point clouds. In addition, we propose a joint loss function for the generator.}
\item{To effectively extract features, we introduce a skip-connected feature extraction module that focuses on highly distorted points. Furthermore, we propose a geometry-guided global feature perception module. This module constructs an extended local neighborhood based on correlations between neighboring patches and incorporates a transformer-based structure to capture long-range global correlations. This improves patch boundary attention, expands the receptive field, and enhances quality of textured areas and patch boundaries.}
\item{Experimental results show that integrating the proposed method into the G-PCC point cloud compression codec significantly improves the rate-distortion (RD) performance in terms of both data fidelity and perceptual quality.}
\end{itemize}

The remainder of this paper is organized as follows. In Section II, we briefly review related work. In Section III, we provide the theoretical motivation of our method and formulate the problem mathematically. In Section IV, we present and analyze the proposed network. Experimental results and conclusions are given in Sections V and VI, respectively.

\section{Related Work}
In this section, we review relevant work on point cloud attribute quality enhancement, focusing on two approaches: traditional methods and deep learning-based methods.
\subsection{Traditional methods for quality enhancement}
Traditional methods for quality enhancement can be roughly divided into two classes: traditional filter-based algorithms and graph signal processing \cite{refa5}-based algorithms.

Traditional filter-based algorithms: Wang et al. \cite{ref13} introduced the Kalman filter for the reconstructed attributes in G-PCC, where the PredLift configuration is activated for attribute compression. This not only improves the quality of the reconstructed attributes but also effectively enhances prediction accuracy. However, the Kalman filter is highly sensitive to signal stationarity, making it primarily effective for chrominance components (i.e., Cb and Cr). To address this, Wiener filter-based methods \cite{ref14,ref15} were subsequently proposed to effectively mitigate distortion accumulation during the coding process and enhance reconstruction quality. While Wiener filters offer certain advantages, their effectiveness is limited by the assumption of linear distortion. 

Graph signal processing algorithms: Yamamoto et al. \cite{refa6} proposed a deblurring algorithm for point cloud attributes, inspired by the multi-Wiener SURE-LET deconvolution method \cite{refa7} used in image processing. This algorithm models blurred textures as graph signals, where the deblurring process begins with Wiener-like filtering, followed by sub-band decomposition and thresholding, to enhance the quality of deblurred point clouds. Dinesh et al. \cite{ref16} proposed two algorithms for 3D point cloud color denoising: one based on graph Laplacian regularization (GLR) and the other on graph total variation (GTV) priors, formulating the denoising problem as a maximum a posteriori estimation and solving it using conjugate gradient for GLR and alternating direction method of multipliers with proximal gradient descent for GTV. Watanabe et al. \cite{ref17} proposed a point cloud color denoising method based on 3D patch similarity, which generates a more robust graph structure by calculating the similarity of 3D patches around connected points, and designed a low-pass filter that automatically selects the frequency response based on the estimated noise level. Later, they \cite{ref18} proposed a fast graph-based denoising method that achieves efficient denoising on large-scale point clouds by employing scan-line neighborhood search for high-speed graph construction, covariance matrix eigenvalues for fast noise estimation, and low-cost filter selection. However, the time complexity and the difficulty of constructing an appropriate graph remain significant challenges for the aforementioned methods.
\vspace{-0.4cm}
\subsection{Deep learning-based methods for quality enhancement}
Deep learning-based approaches for improving the attribute quality of distorted point clouds can be broadly categorized into three methods: graph convolution-based, sparse convolution-based, and projection-based techniques. 

Graph convolution-based methods: Sheng et al. \cite{ref19} proposed a multi-scale graph attention network to remove artifacts in point cloud attributes compressed by G-PCC. They constructed a graph based on geometry coordinates and applied Chebyshev graph convolutions to extract feature information from point cloud attributes. Xing et al. \cite{ref20} introduced a graph-based quality enhancement network that uses geometry information as an auxiliary input. By using graph convolution, the network efficiently extracts local features and can handle point clouds with varying levels of distortion using a single pre-trained model. 

Sparse convolution-based methods: Sparse 3D convolution has proven effective for improving computational efficiency and reducing memory consumption. Liu et al. \cite{ref21} proposed a quality enhancement method for dynamic point clouds based on inter-frame motion prediction. Their method includes an inter-frame motion prediction module with relative position encoding and motion consistency to align the current frame with the reference frame. Ding et al. \cite{refd0} developed a learning-based adaptive in-loop filter for G-PCC. Zhang et al. \cite{ref22} proposed a method to improve the reconstruction quality of both geometry and attributes. This approach first uses linear interpolation to densify the decoded geometry, creating a continuous surface. Then, a Gaussian distance-weighted mapping is used to recolor the enhanced geometry, which is further refined by an attribute enhancement network. Later, they \cite{refd1} proposed a fully data-driven method and a rules-unrolling-based optimization to restore G-PCC compressed point cloud attributes.

Projection-based methods: Gao et al. \cite{ref23} proposed an occupancy-assisted compression artifact reduction method for V-PCC, which takes the occupancy information as a prior knowledge to guide the network to focus on learning the attribute distortions of the occupied regions. Xing et al. \cite{ref24} proposed a U-Net-based quality enhancement method for color attributes of dense 3D point clouds. To mitigate the hole artifacts caused by trisoup-based geometry compression in G-PCC, Tao et al. \cite{ref25} proposed a multi-view projection-based joint geometry and color hole repairing method. In this approach, irregular points are converted into regular pixels on 2D planes through multi-view projection. They also design a multi-view projection-based triangular hole detection scheme based on depth distribution to effectively repair the holes in both geometry and color.

However, the aforementioned methods primarily focus on point-to-point data fidelity and overlook the perceptual quality of human visual system. As a result, these methods often yield suboptimal performance in terms of providing high-quality visual reconstruction. To address this challenge, we propose an attribute quality enhancement algorithm based on optimal transport theory. Our approach aims to find the optimal solution that simultaneously enhances both data fidelity and subjective quality.

\begin{figure}
\centering
\includegraphics[width=3.2in]{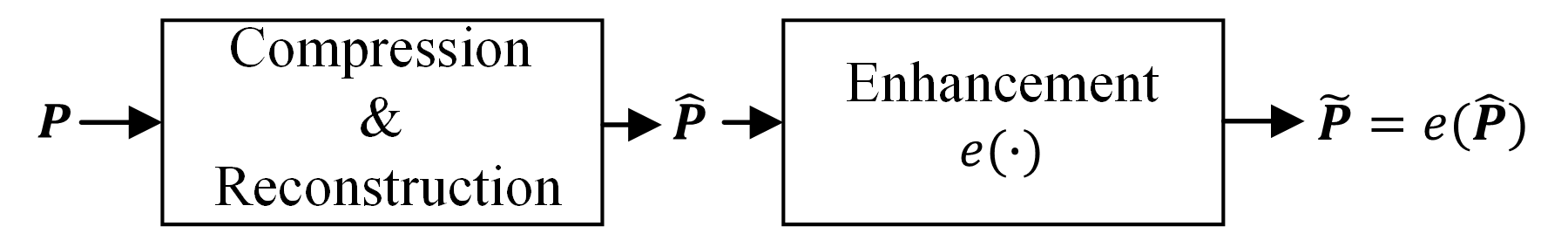}
\caption{Problem description. For an original point cloud \(\bm{P}\) and an initial reconstructed point cloud \(\hat{\bm{P}}\) with severe distortion caused by compression, we aim to obtain an enhanced point cloud \(\tilde{\bm{P}}\) from \(\hat{\bm{P}}\) through an enhancement network \(e(\cdot)\).}
\label{FIG1}
\end{figure}

\section{Problem Formulation}
As shown in Fig. 1, for a reconstructed point cloud \(\hat{\bm{P}} = [\hat{\bm{p}}_1, \hat{\bm{p}}_2, \dots, \hat{\bm{p}}_N]^{\mathsf{T}}  \in \mathbb{R}^{N \times 6}\), where \(\hat{\bm{p}}_i = [\hat{\bm{p}}_i^G, \hat{\bm{p}}_i^A ]\) denotes the \(i^{th}\) point with geometry \(\hat{\bm{p}}_i^G\) and attribute \(\hat{\bm{p}}_i^A\), the enhancement mapping \(\tilde{\bm{P}} := e(\hat{\bm{P}})\) aims to restore \(\hat{\bm{P}}\) to an enhanced version \(\tilde{\bm{P}} = [\tilde{\bm{p}}_1, \tilde{\bm{p}}_2, \dots, \tilde{\bm{p}}_N ]^{\mathsf{T}} \in \mathbb{R}^{N \times 6}\), where \(\tilde{\bm{p}}_i = 
[\tilde{\bm{p}}_i^G, \tilde{\bm{p}}_i^A]\), under the supervision of the original point cloud \(\bm{P} = [\bm{p}_1, \bm{p}_2, \dots, \bm{p}_N ]^{\mathsf{T}}  \in \mathbb{R}^{N \times 6}\), where \(\bm{p}_i = [\bm{p}_i^G, \bm{p}_i^A ]\). An ideal enhancement process should aim to suppress the distortion induced by compression while maximizing the preservation of the original information contained in \(\bm{P}\) through \(\hat{\bm{P}}\). Specifically, this can be achieved by maximizing the mutual information \(I(\bm{P};\hat{\bm{P}})\), which quantifies the amount of shared information between the reconstructed point cloud \(\hat{\bm{P}}\) and the original point cloud \(\bm{P}\). Moreover, perceptual quality, which depends on the deviation from the original signal statistics \cite{ref27,ref28,ref29,ref30}, should also be taken into account. It can be conveniently defined in terms of the deviation between the probability distributions of the enhanced and original signals \cite{ref31}. From the data processing inequality of the Markov chain \(\bm{P}\rightarrow\hat{\bm{P}}\rightarrow e(\hat{\bm{P}})\), we have \(\underset{e}{\max} \ I(e(\hat{\bm{P}}); \bm{P}) \leq I(\hat{\bm{P}}; \bm{P})\). Therefore, the enhancement problem can be expressed as

\begin{equation}
\label{1}
\underset{e}{\max} \ I(e(\hat{\bm{P}}); \bm{P}) \; \text{s.t.} \; d(p_{\bm{P}}, p_{\tilde{\bm{P}}}) \leq \varepsilon,
\end{equation}
where \(d(p_{\bm{P}}, p_{\tilde{\bm{P}}})\) is the deviation between the statistical distributions of \(\bm{P}\) and \(\tilde{\bm{P}}\), i.e., \(p_{\bm{P}}\) and \(p_{\tilde{\bm{P}}}\), and \(\varepsilon\) is an infinitesimal number. Specifically, \(d(\cdot,\cdot)\) can be represented by a divergence (e.g., the Kullback-Leibler divergence) or a distance (e.g., the Wasserstein distance). (1) maximizes the mutual information between the enhanced point cloud \(e(\hat{\bm{P}})\) and the original point cloud  \(\bm{P}\), while constraining the probability distribution  \(p_{\tilde{\bm{P}}}\) of \(e(\hat{\bm{P}})\) to be consistent with the probability distribution \(p_{\bm{P}}\) of \(\bm{P}\). 

The optimal transport problem aims to find the most efficient way to map one distribution of mass to another \cite{ref32,ref33,ref34}. Let \(p_{\bm{P}}\) and \(p_{\hat{\bm{P}}}\) be two sets of probability measures on \(\bm{P}\) and \(\hat{\bm{P}}\), respectively. Let \(c\) be a cost function, where \(c(\hat{\bm{p}}_i, \bm{p}_i)\) represents the cost of transporting \(\hat{\bm{p}}_i \in \hat{\bm{P}}\) to \(\bm{p}_i \in \bm{P}\). The goal of the optimal transport problem is to find the transport plan from \(p_{\hat{\bm{P}}}\) to \(p_{\bm{P}}\) that minimizes this cost.
\vspace{\baselineskip}

\noindent\textbf{Definition 1.} \textit{(Monge’s optimal transport problem): Given two probability measures \(p_{\hat{\bm{P}}}\) and \(p_{\bm{P}}\), the Monge's optimal transport problem seeks a \(p_{\bm{P}}\)-measurable map \(e:\hat{\bm{P}}\rightarrow\bm{P}\)  that attains the infimum}

\begin{equation}
\label{2}
\inf_e  \int_{\hat{\bm{P}}} c(e(\hat{\bm{p}}_i), \bm{p}_i) \, dp_{\hat{\bm{P}}}(\hat{\bm{p}}_i) \; \text{s.t.} \; p_{\bm{p}} = e_\# p_{\hat{\bm{p}}},
\end{equation}
\textit{where \(e_\# p_{\hat{\bm{p}}}\) denotes the transport of \(\hat{\bm{p}}\) by \(e\). A map \(e^\ast\) that achieves this minimum cost, if one exists, is called an optimal transport map.}
\vspace{\baselineskip}

Intuitively, the optimal transport problem finds a transport to turn the mass of \(p_{\hat{\bm{P}}}\) into \(p_{\bm{P}}\) at the minimal cost measured by the cost function \(c\), e.g. \(c(e(\hat{\bm{p}}_i), \bm{p}_i) := \| e(\hat{\bm{p}}_i) - \bm{p}_i \|^\gamma\) with \(\gamma\geq1\).

Based on the above analysis, we model the quality enhancement problem as an implementation of the Monge’s optimal transport problem expressed as follows:

\begin{equation}
\label{3}
\min_e \mathbb{E}_{\hat{\bm{P}} \sim p_{\hat{\bm{P}}}} ( \| \bm{P} - e(\hat{\bm{P}}) \|^\gamma ) \; \text{s.t.} \; p_{\bm{P}} = p_{\tilde{\bm{P}}},
\end{equation}
where \(\mathbb{E}_{\hat{\bm{P}} \sim p_{\hat{\bm{P}}}}(\cdot)\) denotes the mathematical expectation. Specifically, given distorted point cloud \(\hat{\bm{P}}\), it seeks an enhancement map \(e(\cdot)\) that achieves minimum distance transport and high perceptual quality. Specifically, the objective function in (3) ensures the data fidelity of the enhanced point cloud \(e(\hat{\bm{P}})\), by incorporating a minimum distance transport property. This property guarantees that the enhancement map achieves an almost maximal preservation of the information in the original point cloud \(\bm{P}\) that is contained in the reconstructed point cloud \(\hat{\bm{P}}\). The constraint in (3) ensures that the enhanced point cloud \(e(\hat{\bm{P}})\) maintains the same statistical distribution as the original point cloud \(\bm{P}\). This alignment of distributions helps to achieve high perceptual quality, making the enhanced point cloud more consistent with human visual perception.

For ease of implementation, (3) can be relaxed into an unconstrained problem \cite{ref35} as

\begin{equation}
\label{4}
\min_e \mathbb{E}_{\hat{\bm{P}} \sim p_{\hat{\bm{P}}}} ( \| \bm{P} - e(\hat{\bm{P}}) \|^\gamma ) + \lambda d(p_{\bm{P}} , p_{\tilde{\bm{P}}}),
\end{equation}
where \( \lambda > 0 \) is a balance parameter. Though relaxed, (4) has the same solution as (3) when \(d(\cdot,\cdot)\) is the Wasserstein-1 distance, denoted by
\( W_1(\cdot, \cdot) \), where \( \gamma = 1 \) and \( \lambda > 1 \) \cite{ref35}.

\begin{figure*}
\centering
\includegraphics[width=5.6in]{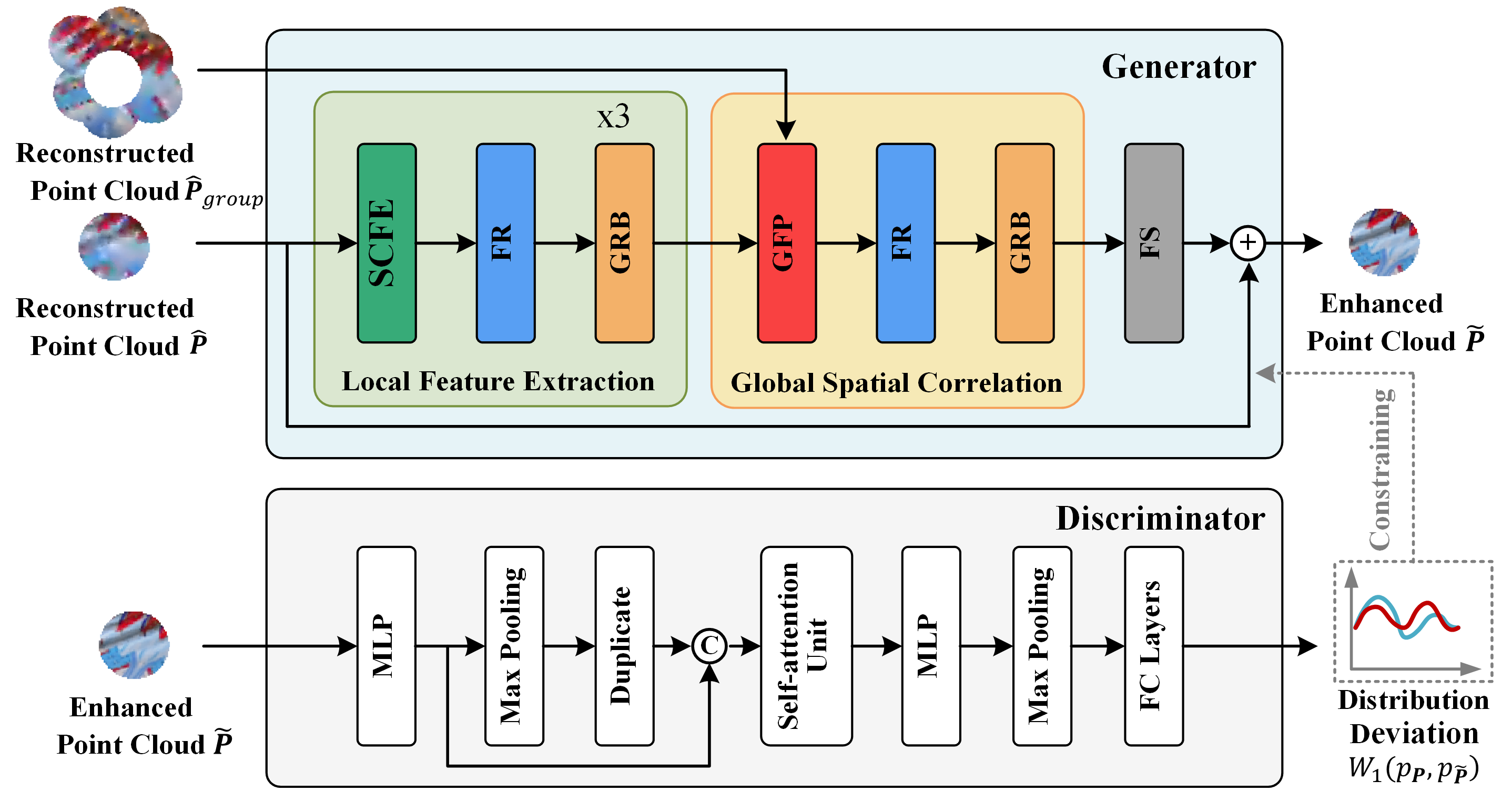}
\caption{Framework of the proposed PCE-GAN, where the generator produces an enhanced point cloud \(\tilde{\bm{P}}\) while the discriminator computes the deviation between the probability distributions of the enhanced point cloud \(\tilde{\bm{P}}\) and the original point cloud \(\bm{P}\), i.e., \(W_1(p_{\bm{P}} , p_{\tilde{\bm{P}}})\).}
\label{FIG2}
\end{figure*}
In summary, our goal is to solve (4), and more specifically to find the optimal enhancement map \(e(\cdot)\). To this end, we represent the enhancement map \(e(\cdot)\) implicitly as a neural network. Specifically, we propose a point cloud quality enhancement network (PCE-GAN) based on the WGAN-GP framework to obtain the enhanced point cloud \(\tilde{\bm{P}}\) whose attribute can be written as

\begin{equation}
\label{5}
\tilde{\bm{P}}^A = e\left( \hat{\bm{P}}^G, \hat{\bm{P}}^A \mid \mathbf{\Theta} \right),
\end{equation}
where \(e(\cdot)\) represents PCE-GAN, \(\mathbf{\Theta}\) represents the learnable parameters, \({\hat{\bm{P}}}^{G}\) and \({\hat{\bm{P}}}^{A}\) are reconstructed geometry and attribute information, respectively. As this paper focuses solely on quality enhancement of attribute, we assume that the geometry is already reconstructed and is available for quality enhancement of attributes.

\section{Proposed Method}
In the proposed PCE-GAN, as shown in Fig. 2, the generator produces an enhanced point cloud \(\tilde{\bm{P}}\) while the discriminator is used to compute the deviation between the probability distributions of \(\tilde{\bm{P}}\) and \(\bm{P}\), i.e., Wasserstein-1 distance \(W_1(p_{\bm{P}} , p_{\tilde{\bm{P}}})\). The generator mainly consists of three units, namely a local feature extraction (LFE) unit, a global spatial correlation (GSC) unit and a feature squeeze (FS) unit, where LFE unit consists of a skip-connected feature extraction (SCFE) module, a feature refinement (FR) module and a graph residual block (GRB), GSC unit consists of a geometry -guided global feature perception (GFP) module, an FR module and a GRB. The discriminator is the same as that of PU-GAN \cite{ref36} expect that the batch norm layers and sigmoid activation are removed, following the strategy of WGAN-GP.

\subsection{Patch Generation and Fusion}
Due to the large number of points in a point cloud, it is challenging to feed it directly into the network for training, therefore we divide the reconstructed point cloud into patches before feeding it into the network. Unless otherwise noted, all mentioned point clouds (e.g., \(\bm{P}\), \(\hat{\bm{P}}\), \(\tilde{\bm{P}})\) in the following statements are patches. First, the reconstructed point cloud is sampled by farthest point sampling (FPS) \cite{ref37} to obtain \(m\) seed points, and for each seed point, the \(k\)-nearest neighbor (KNN) search \cite{ref38} is used to find its \(n-1\) nearest neighbours, so that each seed point and its \(n-1\) nearest neighbours can form a patch \(\hat{\bm{P}}\in\mathbb{R}^{n\times6}\) containing \(n=(N\times ol)/m\) points, where \(ol\) is a parameter that controls the rate of overlap between patches. Second, to alleviate the problem of limited receptive field and lack of correlation between patches, we search for the nearest neighbor patches for each patch to obtain a grouped patch \({\hat{\bm{P}}}_{group}\). This is done as follows: take each seed point as the center point in turn, search for its \({num}_{nei}\) neighboring points among all seed points, and encapsulate the patches formed by its \({num}_{nei}\) neighboring seed points together to obtain the grouped patch \({\hat{\bm{P}}}_{group}\in\mathbb{R}^{{num}_{nei}\times n\times6}\).The patch is then processed by the network to obtain an enhanced patch. These enhanced patches are fused using geometry information to form a complete enhanced point cloud. Because the overlap parameter may result in some points being selected multiple times and others not at all, we average the enhanced values for multiply-selected points and retain the original reconstructed values for unselected points.

\begin{figure}
\centering
\includegraphics[width=2.5in]{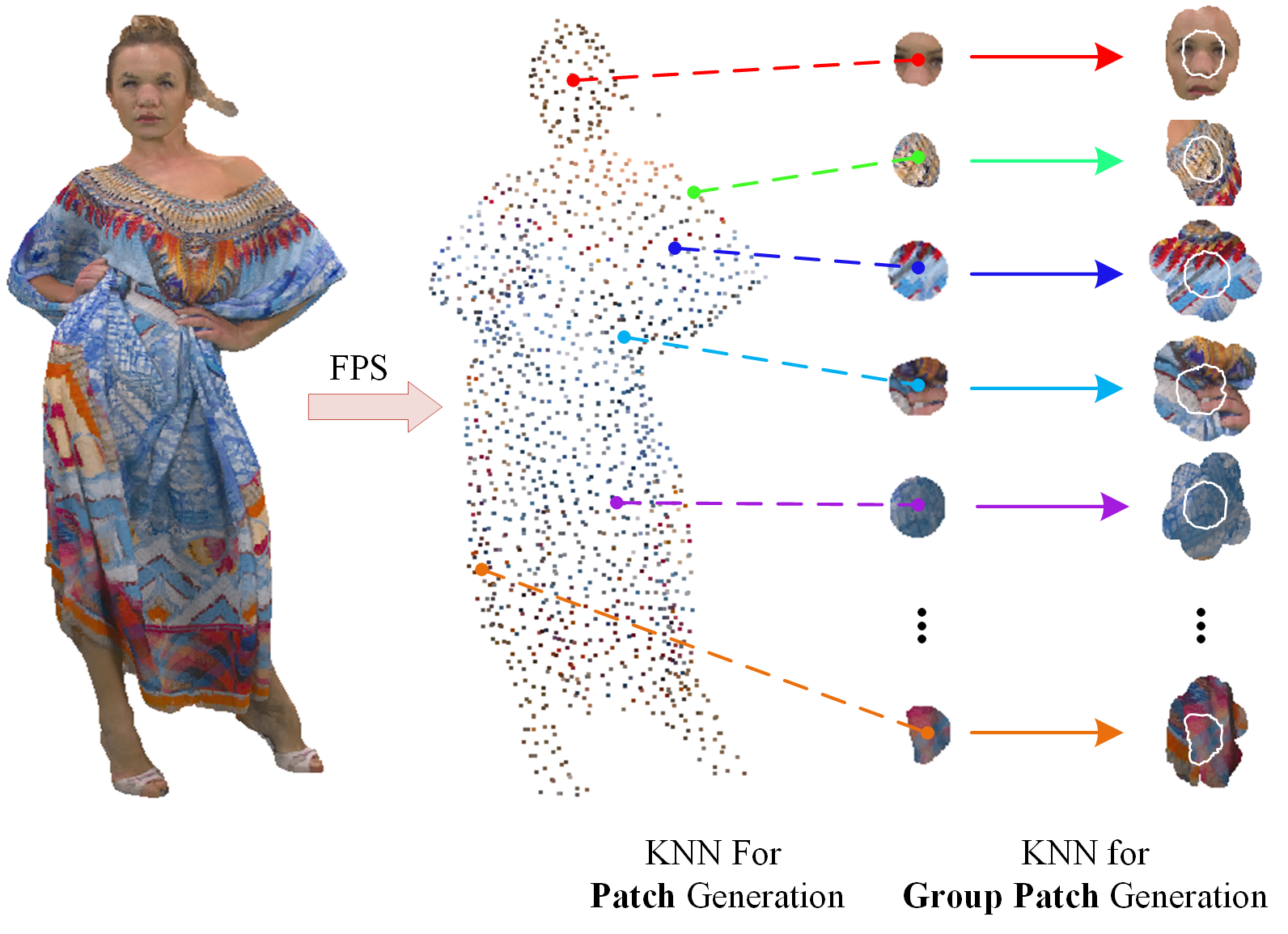}
\caption{Patch generation and grouped patch search.}
\vspace{-0.3cm}
\label{FIG3}
\end{figure}
\begin{figure}
\centering
\includegraphics[width=2.8in]{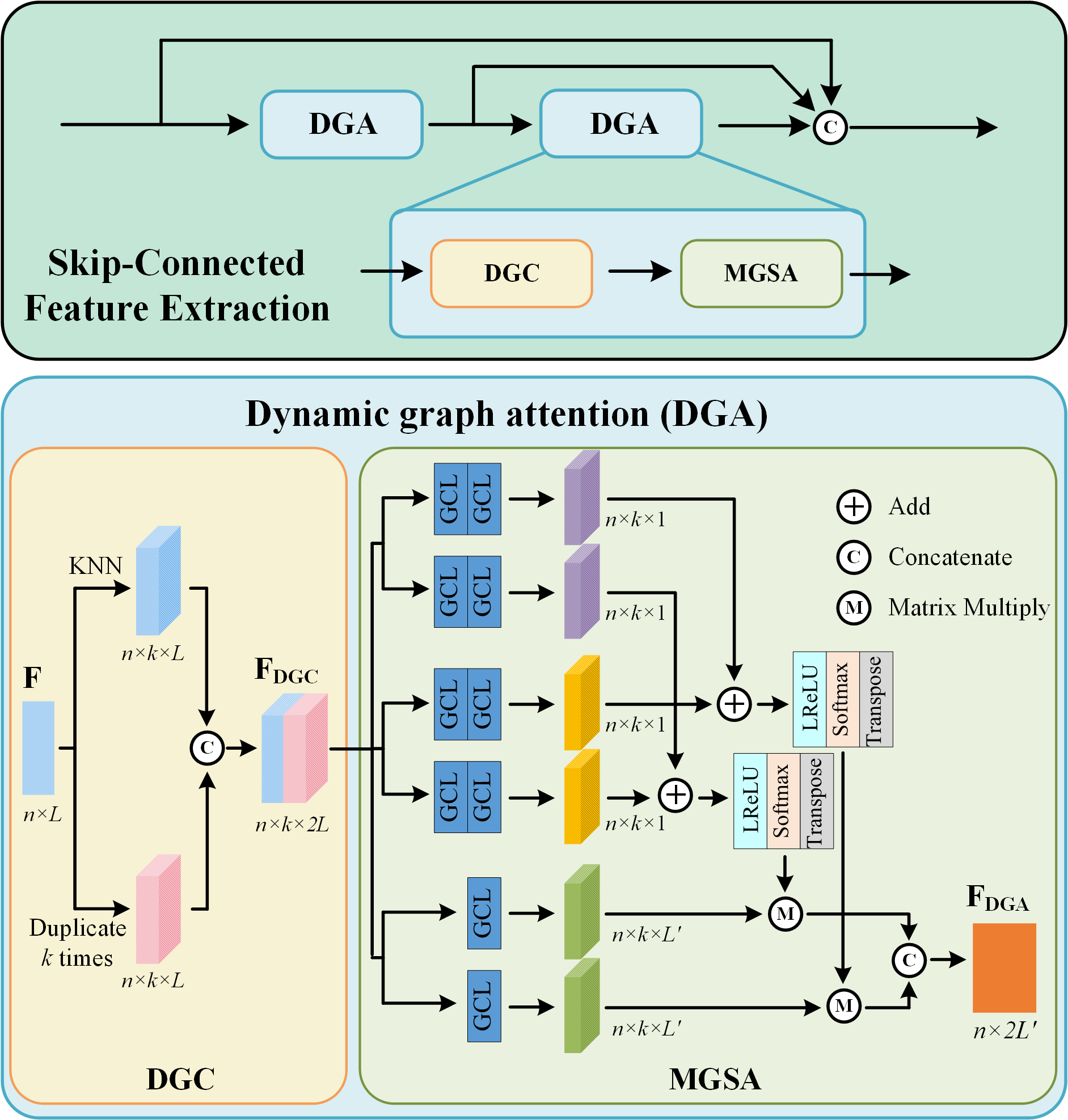}
\caption{Framework of the SCFE module. GCL denotes graph convolution layer.}
\label{FIG4}
\vspace{-0.6cm}
\end{figure}
\subsection{Generator}
The purpose of the generator is to improve a given reconstructed point cloud \(\hat{\bm{P}}\) into an enhanced point cloud \(\tilde{\bm{P}}\). It consists of three units, LFE, GSC and FS, where LFE first extracts the local features of each patch, then, GSC captures the global correlation by using the geometry information of the surrounding neighbor patches to correct the local features, and finally FS maps high-dimensional features from the implicit space into a one-dimensional space. There are five main parts in the generator, namely, SCFE module (in the LFE unit), GFP module (in the GSC unit), GRB (in both the LFE and the GSC unit), FR module (in both the LFE and the GSC unit) and the FS module, which are described in detail as follows.

\begin{figure}
\centering
\includegraphics[width=2.8in]{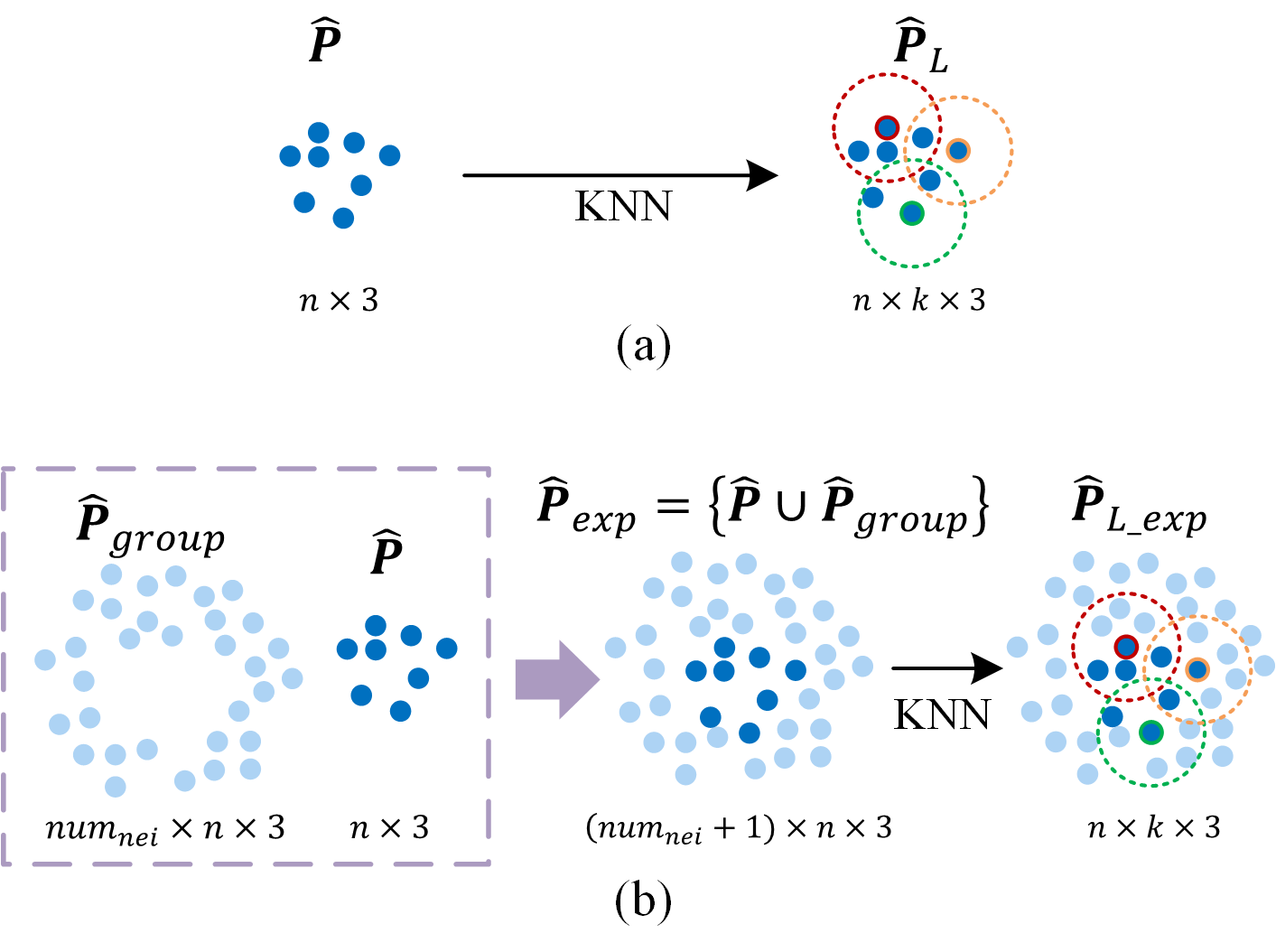}
\caption{Neighborhood construction. (a) Local neighborhood \({\hat{\bm{P}}}_L^i\) construction, (b) expanded local neighborhood \({\hat{\bm{P}}}_{L\_exp}^i\) construction.}
\label{FIG5}
\vspace{-0.6cm}
\end{figure}
\noindent\textit{\textbf{1) SCFE module}}

Points at different locations experience varying levels of distortion after compression, making it crucial to prioritize those with significant distortion. Therefore, we propose a SCFE based on self-attention mechanism \cite{ref39} to extract the features of each point. As shown in Fig. 4, SCFE consists of two cascaded dynamic graph attention (DGA) modules, where DGA consists of a dynamic graph construction (DGC) and a multi-head graph self-attention (MGSA).

DGC extracts rich feature information by dynamically constructing neighborhood graphs in the feature space. Given \(n\) points whose features are represented as \(L\)-dimensional attribute vectors \(\bm{F}=\{f_1, f_2, \dots, f_n\} \in \mathbb{R}^{n \times L}\) and whose 3D geometry coordinates are represented as \(\hat{\bm{P}}^G = [\hat{\bm{p}}_1^G, \hat{\bm{p}}_2^G, \dots, \hat{\bm{p}}_n^G]^{\mathsf{T}} \in \mathbb{R}^{n \times 3}\). DGC first forms each point graph feature \(f_{ij}^\prime\) from a local neighborhood that is computed dynamically via KNN search based on geometry coordinates, i.e., \(f_{ij}^\prime=[f_i\odot{(f}_{ij}-f_i)]\), where \(f_{ij}^\prime\) is the graph feature of the current point and its \(j^{th}\) connected point, \(f_i\) is the feature of the current point. \(f_{ij}\) is the feature of the \(j^{th}\) connected point \(( j=1,2, … ,k)\), and \(\odot\) denotes the concatenation operation. To reduce computational complexity, each point is connected only to its \(k\) nearest neighbors. By combining the features of all the points, the final graph feature \(\bm{F}_{DGC}\in{\mathbb{R}\ }^{n\times k\times2L}\) can be obtained.

Then, MGSA is used to focus on areas with high distortion and complex textures. The input graph feature \(\bm{F}_{DGC}\) is first processed through two graph convolution layers and squeezed to a dimension of \(n\times k\times1\), to extract the most relevant features for each neighboring point, which are then used as the QUERY and KEY vectors. At the same time, \(\bm{F}_{DGC}\) is also projected to a higher-dimensional space \(n\times k\times L^\prime\) through a single graph convolution layer, serving as the VALUE vector. Each graph convolution layer consists of a 2D convolution with a kernel size of \(1\times1\), followed by batch normalization and a Leaky ReLU activation function. Subsequently, the QUERY and KEY vectors are summed, and a Leaky ReLU activation function followed by a softmax function is applied for normalization. The normalized result is then multiplied with the VALUE vector to generate the output feature \(\bm{F}_{DGA}\in\mathbb{R}^{n\times2L^\prime}\). MGSA not only effectively extracts local features from the point cloud but also focuses more on points with severe distortion and complex textures, thereby enhancing the representation of key regions. 

\begin{figure*}
\centering
\includegraphics[width=6.1in]{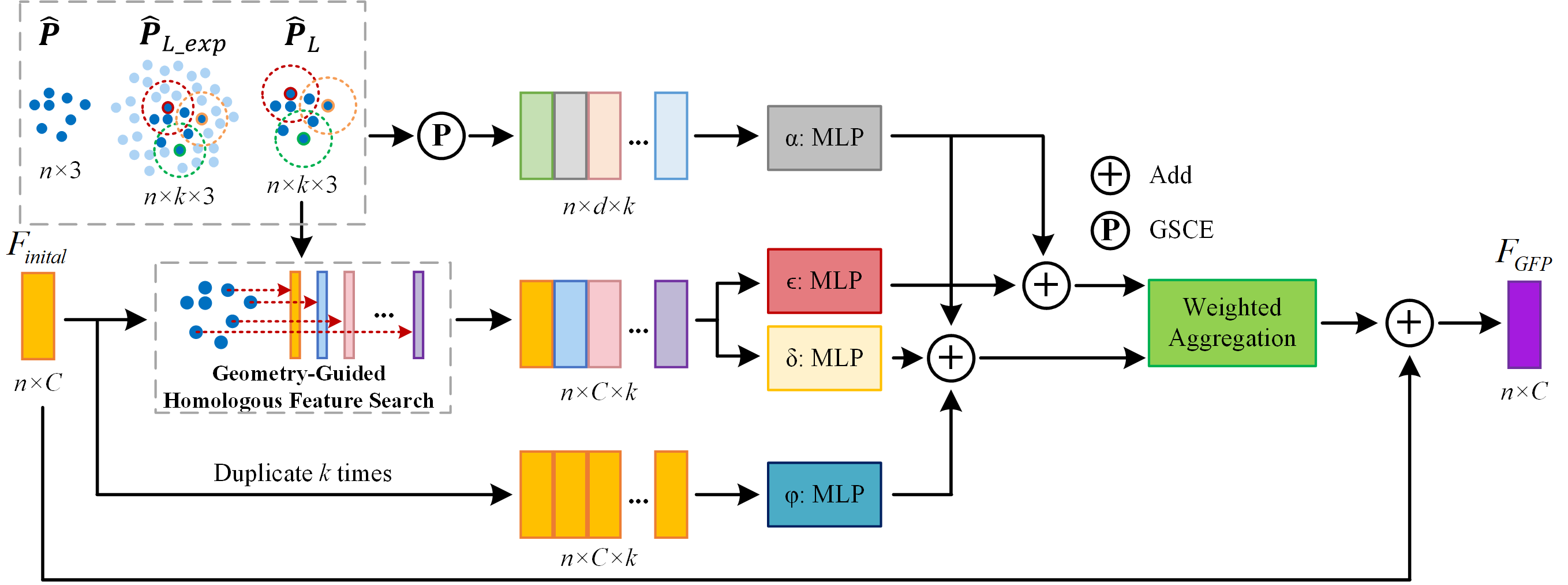}
\caption{Framework of the GFP module.}
\label{FIG6}
\vspace{-0.4cm}
\end{figure*}
\noindent\textit{\textbf{2) GFP module}}

Single-patch based approaches only consider local information, which may result in insufficient attention to points at the patch boundaries, especially for point clouds with complex textures. In such cases, improper treatment of the patch boundaries can lead to a significant reduction in the enhancement effect. However, due to the huge amount of points, it is impractical to feed them directly into the network. Therefore, we propose a GFP module that leverages the correlation between neighboring patches. This method exploits the geometry information from adjacent patches to construct an extended local neighborhood. Additionally, we incorporate a transformer-style structure to capture long-range global correlations. This effectively resolves the issue of insufficient attention to patch boundaries, while also expanding the network's receptive field and enhancing quality of textured areas and patch boundaries. 

Given a patch \(\hat{\bm{P}}\in\mathbb{R}^{n\times6}\) and its grouped patch \({\hat{\bm{P}}}_{group}\in\mathbb{R}^{{num}_{nei}\times n\times6}\), we can obtain the expanded patch \({\hat{\bm{P}}}_{exp}=\{\hat{\bm{P}}\cup{\hat{\bm{P}}}_{group}\}\in\mathbb{R}^{({num}_{nei}+1)\times n\times6}\) and corresponding feature \(\bm{F}_{inital}\in\mathbb{R}^{n\times C}\) of \(\hat{\bm{P}}\), where \(C\) denotes the feature dimension. First, a KNN search is performed on \(\hat{\bm{P}}\) to obtain the \(k\) nearest neighbor points of each point and construct the local neighborhood \({\hat{\bm{P}}}_L^i=\{{\hat{\bm{p}}}_{i1}^G,{\hat{\bm{p}}}_{i2}^G,\ldots,{\hat{\bm{p}}}_{ik}^G\}\), where \({\hat{\bm{p}}}_{ik}^G\in\mathbb{R}^{1\times3}\) denotes the geometry coordinates of the \(k^{th}\) nearest neighbor point of the current point \({\hat{\bm{p}}}_i\). Then, geometry-guided homology feature search is performed on \(\bm{F}_{inital}\) to find the local feature domain map \(\bm{F}_L^i=\{\bm{F}_{i1},\bm{F}_{i2},\ldots,\bm{F}_{ik}\}\), where \(\bm{F}_{ik}\in\mathbb{R}^{1\times c}\) denotes the feature corresponding to the \(k^{th}\) nearest-neighbor point of \({\hat{\bm{p}}}_i\). Second, KNN search is performed again for \(\hat{\bm{P}}\) in \({\hat{\bm{P}}}_{exp}\) to obtain a new expanded local neighborhood \({\hat{\bm{P}}}_{L\_{exp}}^i=\{{{\hat{\bm{p}}}_{i1}^{G}}\prime,{{\hat{\bm{p}}}_{i2}^{G}}\prime,\ldots,{{\hat{\bm{p}}}_{ik}^{G}}\prime\}\) for each point in \(\hat{\bm{P}}\), as shown in Fig. 5. Compared to \({\hat{\bm{P}}}_L^i\), \({\hat{\bm{P}}}_{L\_{exp}}^i\) has the advantage of providing a more complete receptive field for each point on the boundary of \(\hat{\bm{P}}\). To fully leverage the rich geometry information contained in \({\hat{\bm{P}}}_L^i\) and \({\hat{\bm{P}}}_{L\_{exp}}^i\), we design a global spatial correlation-based position encoder (GSCE), 

\begin{eqnarray}
\begin{aligned}
\label{6}
pos_i^k =& \text{GSCE}( \hat{\bm{p}}_i^G \odot \hat{\bm{p}}_{ik}^G \odot {\hat{\bm{p}}_{ik}^G}\prime \odot (\hat{\bm{p}}_i^G - \hat{\bm{p}}_{ik}^G) \\& \odot (\hat{\bm{p}}_i^G - {\hat{\bm{p}}_{ik}^G}\prime) \odot (\hat{\bm{p}}_i^G + \hat{\bm{p}}_{ik}^G - {\hat{\bm{p}}_{ik}^G}\prime )),
\end{aligned}
\end{eqnarray}
\noindent where \({\hat{\bm{p}}}_{i}^{G}\) denotes the geometry coordinates of the current point \({\hat{\bm{p}}}_{i}\), \({\hat{\bm{p}}}_{ik}^{G}\) and \({\hat{\bm{p}}}_{ik}^{G}\prime\) are the geometry coordinates of the \(k^{th}\) nearest neighbors in \({\hat{\bm{P}}}_{L}^{i}\) and \({\hat{\bm{P}}}_{L\_exp}^{i}\), respectively, \({pos}_i^k\) denotes the positional encoding of \({\hat{\bm{p}}}_i\) in the \(k\)-dimension, and \(\odot\) denotes the concatenate operation. As transformers \cite{ref40} can efficiently capture global correlations over long distances through feature similarity, we introduce a transformer-style structure in GFP (Fig. 6) to further correct the features of each point while enhancing attention to the boundary points. Finally, the operation of GFP can be expressed as

\begin{equation}
\begin{aligned}
\bm{F}_{GFP}^{i} = & \bm{F}_{initial}^{i} +\sum_{\bm{F}_{ik} \in \bm{F}_L^{i}} \left( \varphi(\bm{F}_{initial}^{i}) + \delta(\bm{F}_{ik}) + \alpha(pos_i^k) \right) \\& \otimes \left( \epsilon(\bm{F}_{ik}) + \alpha(pos_i^k) \right),
\end{aligned}
\end{equation}
\noindent where \(\bm{F}_{inital}^i\) is the input feature of the GFP corresponding to the initial feature of \({\hat{\bm{p}}}_i\), \(\bm{F}_{ik}\) is the feature corresponding to the \(k^{th}\) nearest neighbors of \({\hat{\bm{p}}}_i\), \(\varphi\left(\cdot\right)\),\(\delta\left(\cdot\right)\), and \(\epsilon(\cdot)\) are the two-layer MLPs, \(\alpha(\cdot)\) is the three-layer MLPs, and \(\otimes\) denotes element-wise multiplication.

\noindent\textit{\textbf{3) GRB }}

GRB introduces graph convolution, residual connection and max pooling, aiming to efficiently extract the graph features and select the most relevant feature to the current point from each channel. As illustrated in Fig. 7, the input features \(\bm{F}_{in}\in \mathbb{R}^{n\times k\times L_{in}}\), where \( L_{in}\) denotes the number of feature channels, is first passed through three 2D convolution layers with a kernel size of \(1\times1\), resulting in intermediate features. Subsequently, the input features and intermediate features are added together, followed by two additional 2D convolution layers and a max pooling layer to get the output features \(\bm{F}_{out}\in \mathbb{R}^{n\times L_{out}}\). At this stage, the neighborhood information for each point can be effectively embedded into \(\bm{F}_{out}\).
\begin{figure}
\centering
\includegraphics[width=2.8in]{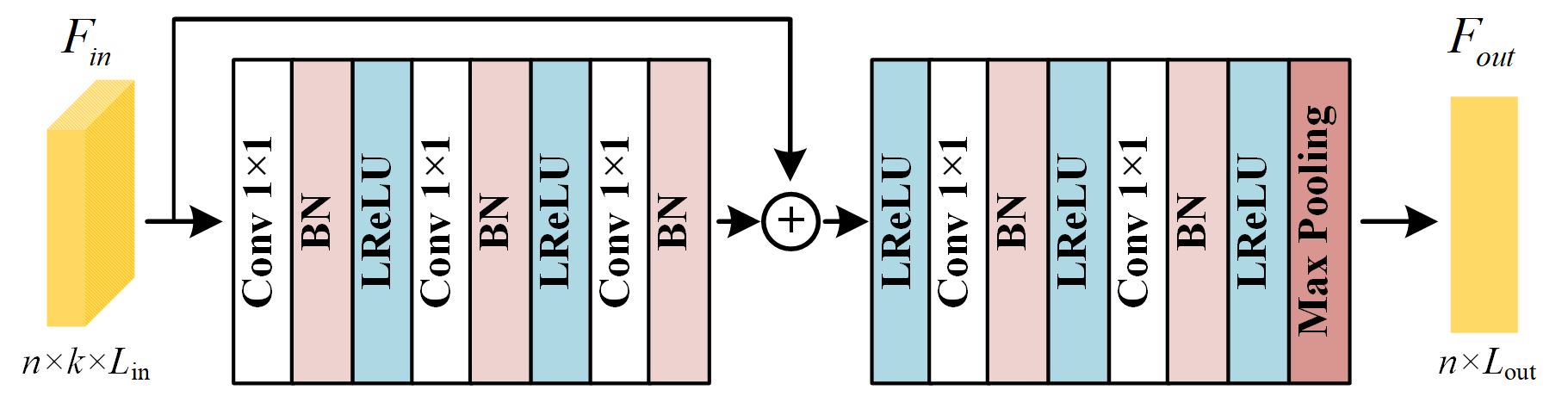}
\caption{Framework of the GRB module.}
\label{FIGGRB}
\end{figure}
\begin{figure}
\centering
\includegraphics[width=2.8in]{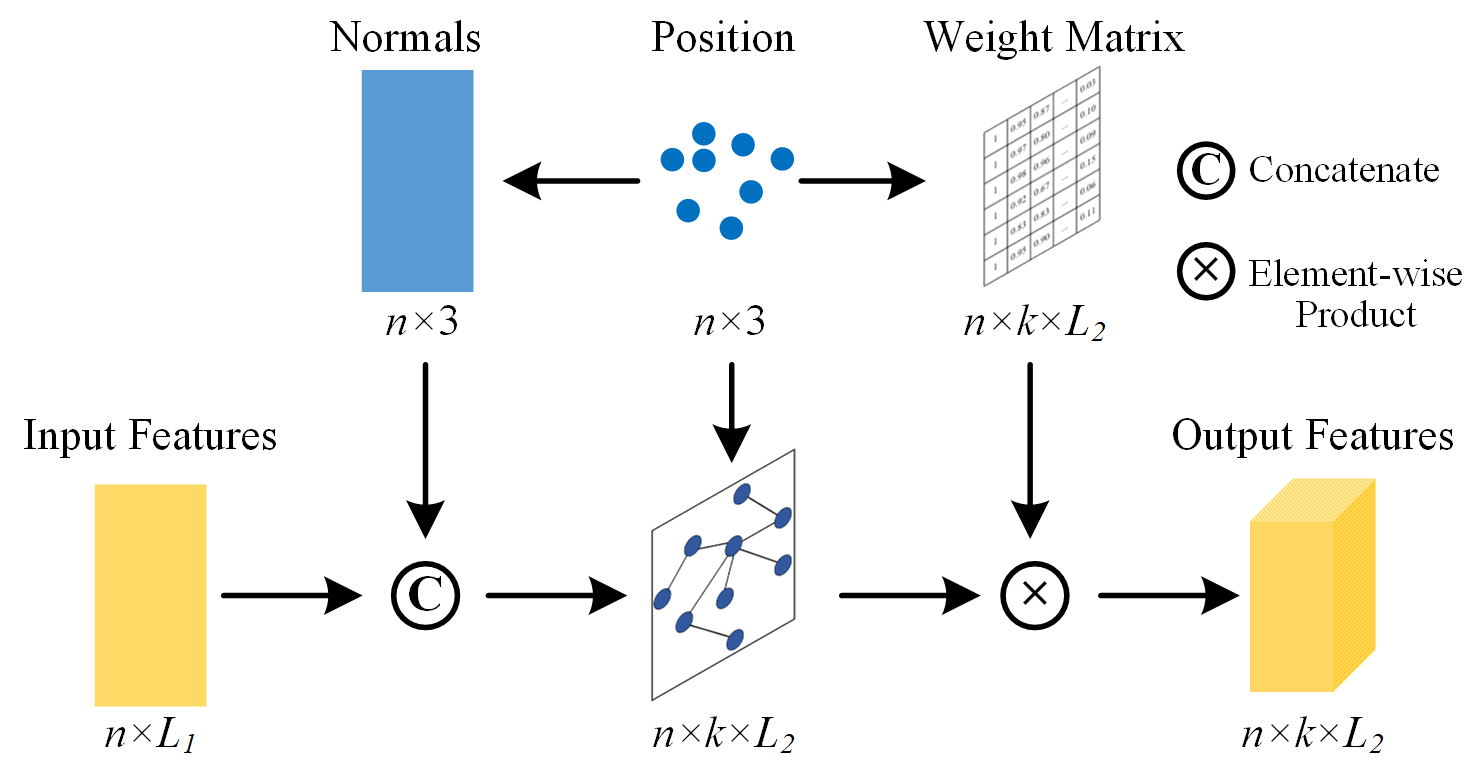}
\caption{Framework of the FR module.}
\vspace{-0.4cm}
\label{FIG7}
\end{figure}

\noindent\textit{\textbf{4) FR module}}

The FR module is borrowed from \cite{ref20}, which effectively exploits the geometry information of each point and its neighborhood to efficiently extract the correlation between points for the purpose of feature correction. As shown in Fig. 8, FR calculates the normal vector of each point and fuses it with the input features. At the same time, it calculates the weight matrix according to the distance between each point and its nearest neighbors in the neighborhood, and then uses the weight matrix to weight the features.

\noindent\textit{\textbf{5) FS module}}

The FS module consists of three graph convolution layers, each comprising a 2D convolution with a kernel size of \(1\times1\), a batch normalization layer, and a Leaky ReLU activation function. The FS module progressively squeezes the input features to a dimension of \(n\times1\), effectively extracting and ultimately capturing the distortion features captured by the network.
\vspace{-0.5cm}
\subsection{Discriminator}
The structure of the discriminator is the same as that of PU-GAN. By following the strategy of WGAN-GP, we remove the batch normalization layers and sigmoid activation function of the discriminator, use the Wasserstein-1 distance as the loss function instead of the cross-entropy loss of the original GAN, and introduce a gradient penalty term to make the discriminator satisfy the K-Lipschitz continuity condition that ensures the feasibility of Wasserstein-1 distance calculation. A detailed derivation of the formulas for how WGAN-GP can compute the Wasserstein-1 distance between the real samples and the generated samples can be found in \cite{ref26,ref41}.
\vspace{-0.5cm}
\subsection{Loss Function}
We propose a joint loss to train the PCE-GAN in an end-to-end manner. Similar to WGAN-GP, the discriminator loss can be written as
\begin{equation}
\small
\label{8}
L_D = \mathbb{E}[D({\bm{P}}^A)] - \mathbb{E}[D(G(\hat{\bm{P}}^A))] + \beta \mathbb{E}[(\|\nabla D(\dddot{\bm{P}}^A)\|_2-1)^2],
\end{equation}

\noindent where \({\bm{P}}^A\) and \(\hat{\bm{P}}^A\) represent the actual attribute and the reconstructed attribute, respectively, \(\dddot{\bm{P}}^A\) is a random interpolation between \({\bm{P}}^A\) and \(G(\hat{\bm{P}}^A)\), \(G(\cdot)\) denotes the generator, \(D(\cdot)\) denotes the discriminator, \(\mathbb{E}[\cdot]\) denotes the mathematical expectation, \(\nabla\) denotes the gradient operator, and \(\beta\) is a balancing parameter.

Based on the analysis in Section III, we introduce a fidelity constraint between the enhanced point cloud \(G(\hat{\bm{P}}^A)\) and the original point cloud \({\bm{P}}^A\), which is modelled as a minimum transport distance using the root mean square error (RMSE),

\begin{equation}
\label{9}
L_{{RMSE}} = \sqrt{\frac{1}{n} \sum_{i=1}^{n}\| G(\hat{\bm{p}}_i^A) - \bm{p}_i^A \|^2},
\end{equation}

\noindent Therefore, the generator loss can be expressed as a joint loss of fidelity (\(L_{RMSE}\)) between the enhanced point cloud \(G(\hat{\bm{P}}^A)\) and the original point cloud \({\bm{P}}^A\) as well as the difference in their probability distributions (\(\mathbb{E}[{D(G(\hat{\bm{P}}}^A))]\)), i.e.,

\begin{equation}
\label{10}
L_G = \omega L_{{RMSE}} - \mathbb{E}[D(G(\hat{\bm{P}}^A))].
\end{equation}

\noindent where \(\omega\) is an expansion parameter applied to enlarge \(L_{RMSE}\).

During training, the discriminator is first trained using the generator’s initial output as input. The generator is then refined with fixed discriminator parameters. This iterative process continues until a predefined number of epochs is reached, at which point the enhanced point cloud can be obtained from the generator.

\section{Experimental Results and Analysis}
We conducted extensive experiments to assess the enhanced point clouds in both objective and subjective quality. We also compared the coding efficiency before and after integrating the proposed method into the G-PCC-based compression system. Furthermore, we analyzed how each component of PCE-GAN contributed to the overall performance. 
\vspace{-0.6cm}
\subsection{Experimental Setup}
\noindent\textit{\textbf{1) Datasets}}

We used point clouds with color attributes from the Waterloo point cloud dataset (WPCD) \cite{ref42} as the training dataset. During training, we set the number of points \(n\) in a patch to 2048, the overlapping ratio \(ol\) to 2, and \({num}_{nei}\) to 6. We encoded the point clouds using G-PCC Test Model Category 13 version 26.0 (TMC13v26) \cite{ref43}, applying both the PredLift and RAHT configurations to generate two separate training datasets. The encoding was conducted under the Common Test Condition (CTC) C1 \cite{ref44}, which involves lossless geometry compression and lossy attribute compression. In total, we collected 27860 patches in each training dataset. We trained two models: one for PredLift and the other for RAHT. 
\begin{figure}
\centering
\includegraphics[width=2.6in]{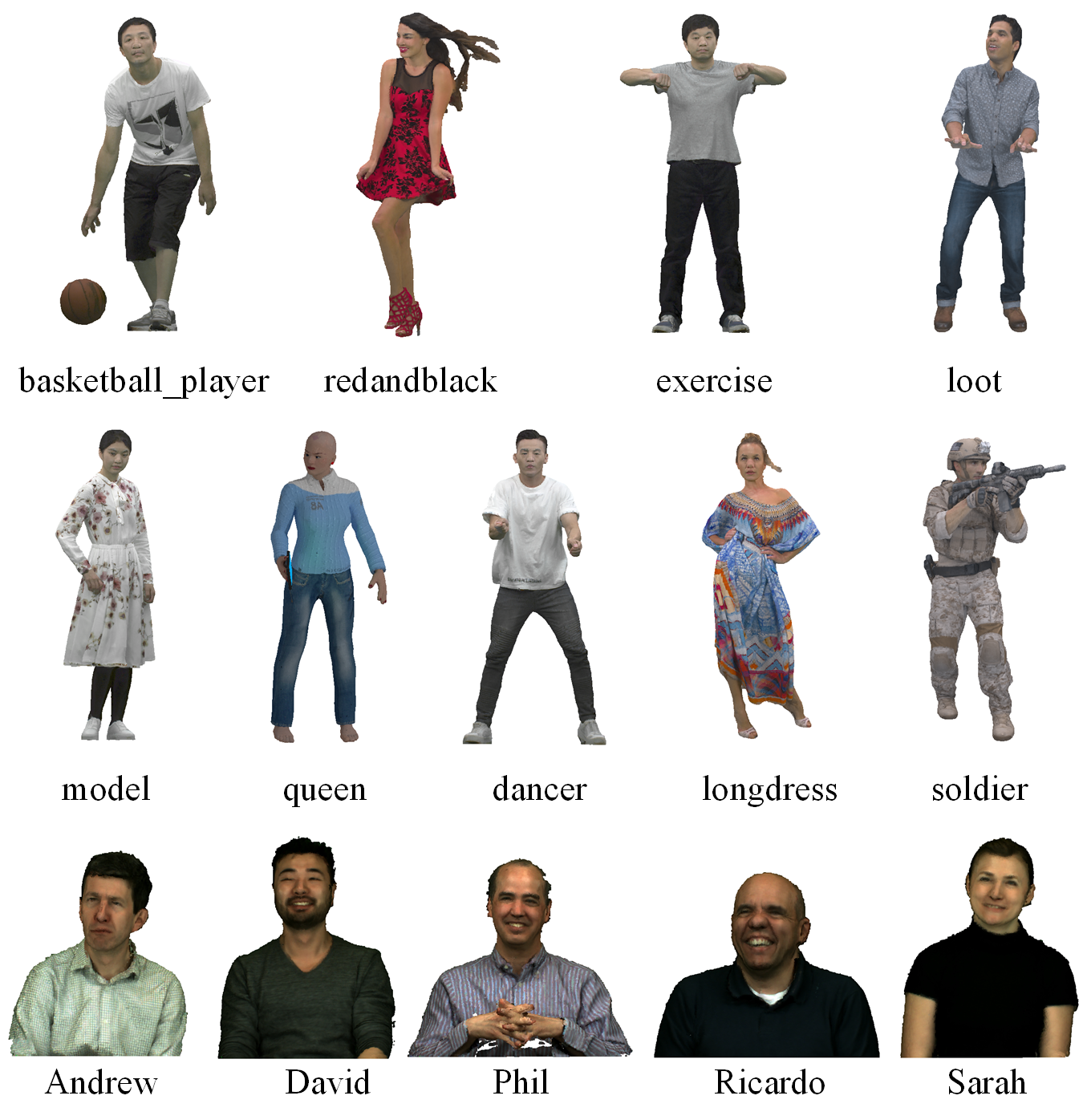}
\caption{Testing dataset.}
\vspace{-0.5cm}
\label{FIG9}
\end{figure}

We tested the trained models on 14 point clouds from the MPEG dataset \cite{ref45} (Fig.9). Each point cloud was compressed using TMC13v26 with quantization parameters (QPs) 51, 46, 40, 34, 28, 22, corresponding to the six bitrates, R01, R02, . . ., R06.

\noindent\textit{\textbf{2) Implementation Details}}

We trained the proposed PCE-GAN for 30 epochs with a batch size of 5. We used the Adam optimizer \cite{ref46}, with learning rates of 0.0001 for the generator and 0.000001 for the discriminator. Moreover, we set \(k=20\), \(\beta=10\), and \(\omega=60\). We implemented the proposed method on an NVIDIA GeForce RTX4090 GPU, using PyTorch v1.12. We trained three models, one for each color attribute component (Y, Cb, and Cr). Each component was processed independently. 
\vspace{-0.5cm}
\subsection{Data Fidelity Evaluation}
We used the peak signal-to-noise ratio (PSNR) to evaluate the data fidelity. Furthermore, to evaluate the coding efficiency before and after integrating the proposed method into the coding system, we used the BD-PSNR, \(\Delta\)PSNR, and BD-rate metrics \cite{ref47}. The BD-rate measures the average bitrate reduction in bits per input point (bpip) at the same PSNR, while the BD-PSNR measures the average PSNR increment in dB at the same bitrate. A negative BD-rate and a positive BD-PSNR indicate better RD performance. In addition to calculating the PSNR for all the color components, we also used YCbCr-PSNR \cite{ref48} to evaluate the overall color quality gains brought by the proposed method. On the other hand, \(\Delta\)PSNR measures the PSNR difference between the proposed method and the anchor at a single bitrate. A positive \(\Delta\)PSNR value indicates that the proposed method achieved a higher PSNR, and thus better performance. 

Table I shows the BD-PSNRs and BD-rates achieved by PCE-GAN when PredLift configuration was used. We can see that PCE-GAN achieved average BD-PSNRs of 0.75 dB, 0.32 dB, and 0.49 dB for the Y, Cb, and Cr components, respectively, corresponding to -19.7\%, -13.4\%, and -17.3\% BD-rates, respectively. The largest PSNR gains were notably high, reaching 1.66 dB for the Luma component of the \textit{soldier} at R05, 1.07 dB for the Cb component of the \textit{longdress} at R06, and 1.75 dB for the Cr component of the \textit{redandblack} at R05. Additionally, we observed that PCE-GAN performed better on point clouds with complex textures, resulting in a significant improvement in PSNR. This indicates that PCE-GAN was better suited for processing point clouds with intricate textures.
\begin{table}[ht]
\centering
\caption{RATE-DISTORTION COMPARISON WITH TMC13V26 PREDLIFT}
\label{tab:my_label1}  
\resizebox{8.55cm}{!}{
\begin{tabular}{lcccccccc}
\toprule
\multirow{2}{*}{\textbf{Sequence}} & \multicolumn{4}{c}{\textbf{BD-PSNR (dB)}} & \multicolumn{4}{c}{\textbf{BD-rate (\%)}} \\ 
 & \textbf{Y} & \textbf{Cb} & \textbf{Cr} & \textbf{YCbCr} & \textbf{Y} & \textbf{Cb} & \textbf{Cr} & \textbf{YCbCr} \\ \hline
       basketball & 0.654 & 0.213 & 0.384 & 0.610 & -20.2 & -13.0 & -18.3 & -19.6 \\ 
        dancer & 0.713 & 0.223 & 0.393 & 0.662 & -20.6 & -13.2 & -16.9 & -19.9 \\ 
        exercise & 0.366 & 0.156 & 0.261 & 0.347 & -15.1 & -12.2 & -19.4 & -15.2 \\ 
        longdress & 0.883 & 0.496 & 0.864 & 0.857 & -18.4 & -12.9 & -20.5 & -18.2 \\ 
        loot & 0.713 & 0.470 & 0.586 & 0.690 & -18.8 & -17.3 & -20.2 & -18.8 \\ 
        model & 0.705 & 0.239 & 0.552 & 0.666 & -18.4 & -11.0 & -19.4 & -18.0 \\ 
        queen & 0.659 & \textbf{0.747} & 0.804 & 0.674 & -16.9 & \textbf{-21.8} & -22.8 & -17.5 \\ 
        redandblack & 0.719 & 0.284 & \textbf{1.230} & 0.723 & -18.8 & -11.1 & \textbf{-25.8} & -18.7 \\ 
        soldier & 0.973 & 0.342 & 0.424 & 0.899 & -21.8 & -12.5 & -14.6 & -20.8 \\ 
        Andrew & 1.029 & 0.186 & 0.186 & 0.923 & -25.2 & -9.8 & -9.4 & -23.2 \\ 
        David & 0.474 & 0.237 & 0.283 & 0.447 & -14.5 & -14.7 & -15.9 & -14.6 \\ 
        Phil & \textbf{1.044} & 0.176 & 0.295 & \textbf{0.943} & \textbf{-26.0} & -7.6 & -10.9 & \textbf{-23.9} \\ 
        Ricardo & 0.690 & 0.329 & 0.334 & 0.646 & -19.7 & -17.8 & -16.0 & -19.4 \\ 
        Sarah & 0.830 & 0.332 & 0.288 & 0.765 & -22.0 & -13.1 & -12.8 & -20.9 \\ \hline
\textbf{Average} & \textbf{0.747} & \textbf{0.316} & \textbf{0.492} & \textbf{0.704} & \textbf{-19.7} & \textbf{-13.4} & \textbf{-17.3} & \textbf{-19.2} \\ \bottomrule
\end{tabular}}
\end{table}
\begin{table}[ht]
\centering
\caption{RATE-DISTORTION COMPARISON WITH TMC13V26 RAHT}
\label{tab:my_label2}  
\resizebox{8.55cm}{!}{
\begin{tabular}{lcccccccc}
\toprule
\multirow{2}{*}{\textbf{Sequence}} & \multicolumn{4}{c}{\textbf{BD-PSNR (dB)}} & \multicolumn{4}{c}{\textbf{BD-rate (\%)}} \\ 
 & \textbf{Y} & \textbf{Cb} & \textbf{Cr} & \textbf{YCbCr} & \textbf{Y} & \textbf{Cb} & \textbf{Cr} & \textbf{YCbCr} \\ \hline
        basketball & 0.688 & 0.231 & 0.581 & 0.653 & -19.9 & -15.4 & -26.7 & -20.0 \\ 
        dancer & 0.735 & 0.239 & 0.612 & 0.696 & \textbf{-20.1} & -14.3 & -24.5 & -20.0 \\ 
        exercise & 0.380 & 0.186 & 0.453 & 0.372 & -15.1 & -15.8 & \textbf{-31.3} & -16.2 \\ 
        longdress & 0.584 & 0.508 & 0.722 & 0.588 & -14.3 & -15.0 & -18.9 & -14.6 \\ 
        loot & 0.529 & 0.569 & 0.839 & 0.551 & -15.9 & -23.3 & -30.6 & -17.3 \\ 
        model & 0.677 & 0.275 & 0.678 & 0.652 & -17.8 & -13.9 & -23.8 & -18.0 \\ 
        queen & 0.739 & \textbf{1.020} & 0.913 & 0.768 & -19.3 & \textbf{-28.9} & -24.6 & \textbf{-20.3} \\ 
        redandblack & 0.625 & 0.382 & \textbf{0.953} & 0.630 & -17.5 & -16.5 & -20.6 & -17.6 \\ 
        soldier & 0.700 & 0.439 & 0.558 & 0.675 & -18.1 & -20.0 & -22.8 & -18.5 \\ 
        Andrew & 0.576 & 0.173 & 0.187 & 0.527 & -16.3 & -11.2 & -11.2 & -15.7 \\ 
        David & 0.447 & 0.305 & 0.426 & 0.437 & -13.0 & -15.2 & -21.1 & -13.7 \\ 
        Phil & 0.706 & 0.201 & 0.299 & 0.649 & -18.0 & -8.8 & -10.9 & -17.0 \\ 
        Ricardo & 0.573 & 0.397 & 0.495 & 0.557 & -14.7 & -18.2 & -20.0 & -15.3 \\ 
        Sarah & \textbf{0.798} & 0.539 & 0.599 & \textbf{0.769} & -17.9 & -17.2 & -20.4 & -18.0 \\ \hline
\textbf{Average} & \textbf{0.625} & \textbf{0.390} & \textbf{0.594} & \textbf{0.609} & \textbf{-17.9} & \textbf{-19.0} & \textbf{-24.1} & \textbf{-18.3} \\ \bottomrule
\end{tabular}}
\end{table}

\begin{figure*}
\centering
\includegraphics[width=6.6in]{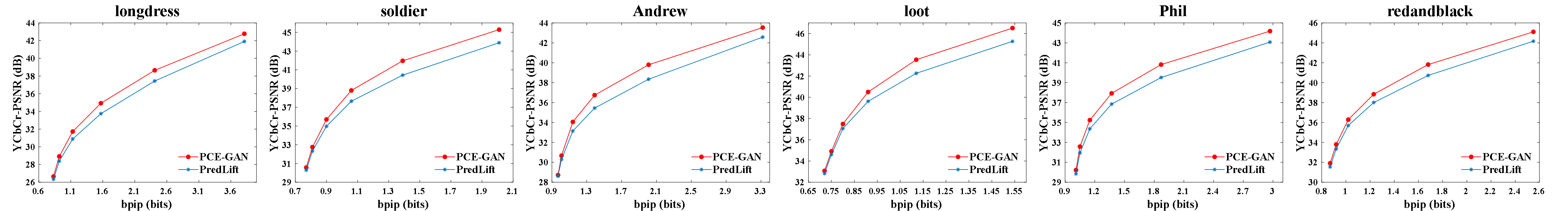}
\caption{Comparison of rate-PSNR curves between PCE-GAN and G-PCC with PredLift configuration.}
\label{FIG10}
\end{figure*}
\begin{figure*}
\centering
\includegraphics[width=6.6in]{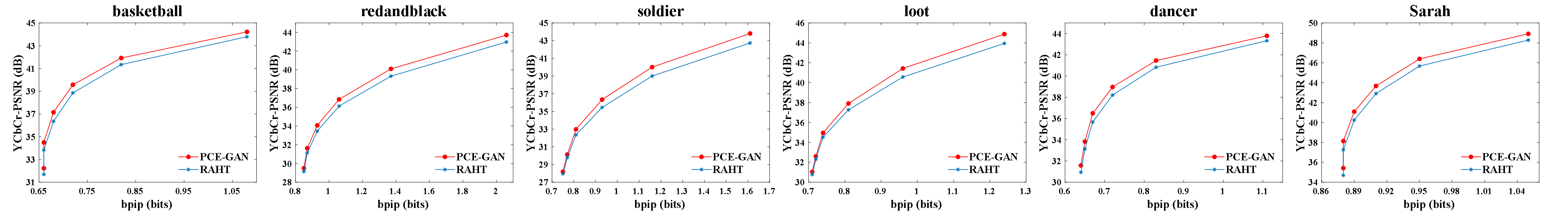}
\caption{Comparison of rate-PSNR curves between PCE-GAN and G-PCC with RAHT configuration.}
\label{FIG11}
\end{figure*}
Similarly, Table II shows the BD-PSNRs and BD-rates achieved by PCE-GAN when the RAHT configuration was used. PCE-GAN achieved average BD-PSNRs of 0.63 dB, 0.39 dB, and 0.59 dB for the Y, Cb, and Cr components, respectively, corresponding to -17.9\%, -19.0\%, and -24.1\% BD-rates. The largest PSNR gains were 1.08 dB (Luma component of \textit{soldier} at R06), 1.17 dB (Cb component of \textit{queen} at R03), and 1.34 dB (Cr component of \textit{redandblack} at R04). 

The rate-PSNR curves before and after integrating PCE-GAN into G-PCC are compared in Fig. 10 (PredLift configuration) and Fig. 11 (RAHT configuration). The results show that the proposed method significantly improved coding efficiency, particularly at medium and high bitrates.

\subsection{Perceptual Quality Evaluation}
To evaluate subjective quality, we used one of the most effective full-reference point cloud quality assessment methods: information content weighted structural similarity (\(\text{IWSSIM}_{\text{P}}\)) \cite{ref49}. Inspired by the classical perceptual image quality assessment algorithm IWSSIM \cite{ref50}, this method objectively compares perceptual quality in alignment with the human visual system. We calculated the perceptual quality improvement provided by PCE-GAN using \(\Delta\text{IWSSIM}_{\text{P}}\), which measures the improvement in \(\text{IWSSIM}_{\text{P}}\) at the same bitrate. A positive \(\Delta\text{IWSSIM}_{\text{P}}\) indicates improved perceptual quality, with larger values representing greater improvement. To measure the average perceptual quality difference across the entire bitrate range, we also employ \(\text{BD-IWSSIM}_{\text{P}}\), which is obtained by integrating and averaging \(\Delta\text{IWSSIM}_{\text{P}}\) over multiple bitrates. A positive \(\text{BD-IWSSIM}_{\text{P}}\) indicates improved perceptual quality, with larger values representing greater improvement.
\begin{table}
\centering
\caption{PERCEPTUAL QUALITY COMPARISON WITH TMC13V26 PREDLIFT}
\label{tab:my_label3}  
\resizebox{8.4cm}{!}{
\begin{tabular}{lcccccc}
\toprule
\multirow{2}{*}{\textbf{Sequence}} & \multicolumn{6}{c}{\(\Delta\textbf{IWSSIM}_{\textbf{P}}\:(\textbf{10}^{\textbf{-3}})\:\uparrow \)} \\ 
 & \textbf{R01} & \textbf{R02} & \textbf{R03} & \textbf{R04} & \textbf{R05} & \textbf{R06}  \\ \hline
        basketball & 2.3519 & 1.7977 & 1.1808 & 0.5366 & 0.2396 & 0.0968 \\ 
        dancer & 2.8096 & 1.9873 & 1.4033 & 0.6775 & 0.3137 & 0.1236 \\ 
        exercise & 2.2941 & 1.5370 & 0.8051 & 0.4038 & 0.2554 & 0.1446 \\ 
        longdress & 3.2822 & \textbf{3.2949} & \textbf{2.4144} & \textbf{1.5192} & 0.6423 & 0.1586 \\ 
        loot & 2.3878 & 1.7470 & 1.0040 & 0.9279 & 0.6396 & 0.2943 \\ 
        model & \textbf{3.6936} & 2.6447 & 1.5811 & 0.7465 & 0.3199 & 0.1059 \\ 
        queen & 1.9461 & 1.6094 & 0.7728 & 0.5360 & 0.3453 & 0.1578 \\ 
        redandblack & 2.4818 & 1.4812 & 0.8831 & 0.7264 & 0.4889 & 0.1710 \\ 
        soldier & 2.2954 & 2.4041 & 1.9500 & 1.3294 & \textbf{0.8364} & \textbf{0.3255} \\ 
        Andrew & 1.1455 & 1.5559 & 1.9542 & 1.2999 & 0.7197 & 0.1696 \\ 
        David & 0.5194 & 0.3545 & 0.2262 & 0.1511 & 0.0825 & 0.0476 \\ 
        Phil & 1.9337 & 2.4817 & 1.4902 & 0.7294 & 0.4163 & 0.1471 \\ 
        Ricardo & 0.6132 & 0.5717 & 0.3206 & 0.1556 & 0.0817 & 0.0231 \\ 
        Sarah & 1.3005 & 0.7632 & 0.4502 & 0.2196 & 0.1083 & 0.0529 \\ \hline
\multirow{2}{*}{\textbf{Average}} & \textbf{2.0753} & \textbf{1.7307} & \textbf{1.1740} & \textbf{0.7114} & \textbf{0.3921} & \textbf{0.1442} \\ \cline{2-7}
& \multicolumn{6}{c}{\textbf{1.0380}} \\ \bottomrule
\end{tabular}}
\end{table}

Tables III and IV show the \(\Delta\text{IWSSIM}_{\text{P}}\) of each bitrate under PredLift and RAHT configurations, respectively. Table III shows that PCE-GAN achieved an average \(\Delta\text{IWSSIM}_{\text{P}}\) of 1.0380. The perceptual quality gain of PCE-GAN decreased from low to high bitrates, indicating that PCE-GAN aligned well with practical application requirements. For the human visual system, the enhancement of perceptual quality at low bitrates is of greater importance than at high bitrates. Similarly, Table IV shows that PCE-GAN achieved an average \(\Delta\text{IWSSIM}_{\text{P}}\) of 2.4746. The results followed the same trend as those in Table III, demonstrating that PCE-GAN achieved a more significant enhancement in perceptual quality at low bitrates. Moreover, PCE-GAN provided a more pronounced improvement in the perceptual quality of point clouds compressed by RAHT. \begin{table}[ht]
\centering
\caption{PERCEPTUAL QUALITY COMPARISON WITH TMC13V26 RAHT}
\label{tab:my_label4}  
\resizebox{8.4cm}{!}{
\begin{tabular}{lcccccc}
\toprule
\multirow{2}{*}{\textbf{Sequence}} & \multicolumn{6}{c}{\(\Delta\textbf{IWSSIM}_{\textbf{P}}\:(\textbf{10}^{\textbf{-3}})\:\uparrow \)} \\ 
 & \textbf{R01} & \textbf{R02} & \textbf{R03} & \textbf{R04} & \textbf{R05} & \textbf{R06}  \\ \hline
        basketball & 7.1750 & 5.0207 & 2.9410 & 1.3613 & 0.5245 & 0.1932 \\ 
        dancer & \textbf{8.2973} & 5.0566 & 2.9435 & 1.5672 & 0.6249 & 0.2188 \\ 
        exercise & 6.5278 & 4.4434 & 2.3425 & 1.1162 & 0.4195 & 0.1608 \\ 
        longdress & 4.2348 & 6.2972 & 4.1966 & 2.0671 & 0.6943 & 0.1925 \\ 
        loot & 6.0610 & 4.0350 & 2.5101 & 1.5158 & 0.7926 & 0.3340 \\ 
        model & 8.2069 & \textbf{6.6325} & 3.5557 & 1.6706 & 0.5858 & 0.1841 \\ 
        queen & 6.2042 & 4.7378 & 2.4650 & 1.3058 & 0.6327 & 0.2323 \\ 
        redandblack & 5.4360 & 4.1323 & 2.5099 & 1.2869 & 0.6078 & 0.2193 \\ 
        soldier & 4.6914 & 4.8690 & \textbf{4.4702} & \textbf{2.5104} & \textbf{0.9979} & \textbf{0.3738} \\ 
        Andrew & 5.1969 & 3.1321 & 4.2865 & 1.8798 & 0.5789 & 0.1455 \\ 
        David & 2.9777 & 1.4387 & 1.7293 & 1.0951 & 0.4579 & 0.1906 \\ 
        Phil & 4.0522 & 5.7332 & 3.3198 & 1.3135 & 0.4041 & 0.1272 \\ 
        Ricardo & 2.2165 & 1.6217 & 1.2986 & 0.6176 & 0.2583 & 0.0692 \\ 
        Sarah & 5.7410 & 3.1558 & 1.4856 & 0.6189 & 0.2553 & 0.0891 \\ \hline
\multirow{2}{*}{\textbf{Average}} & \textbf{5.5013} & \textbf{4.3076} & \textbf{2.8610} & \textbf{1.4233} & \textbf{0.5596} & \textbf{0.1950} \\ \cline{2-7}
& \multicolumn{6}{c}{\textbf{2.4746}} \\ \bottomrule
\end{tabular}}
\end{table}This is because point clouds compressed by RAHT have a lower average bitrate (corresponding to relatively lower reconstruction quality) than point clouds compressed by PredLift.
\begin{figure*}
\centering
\includegraphics[width=5in]{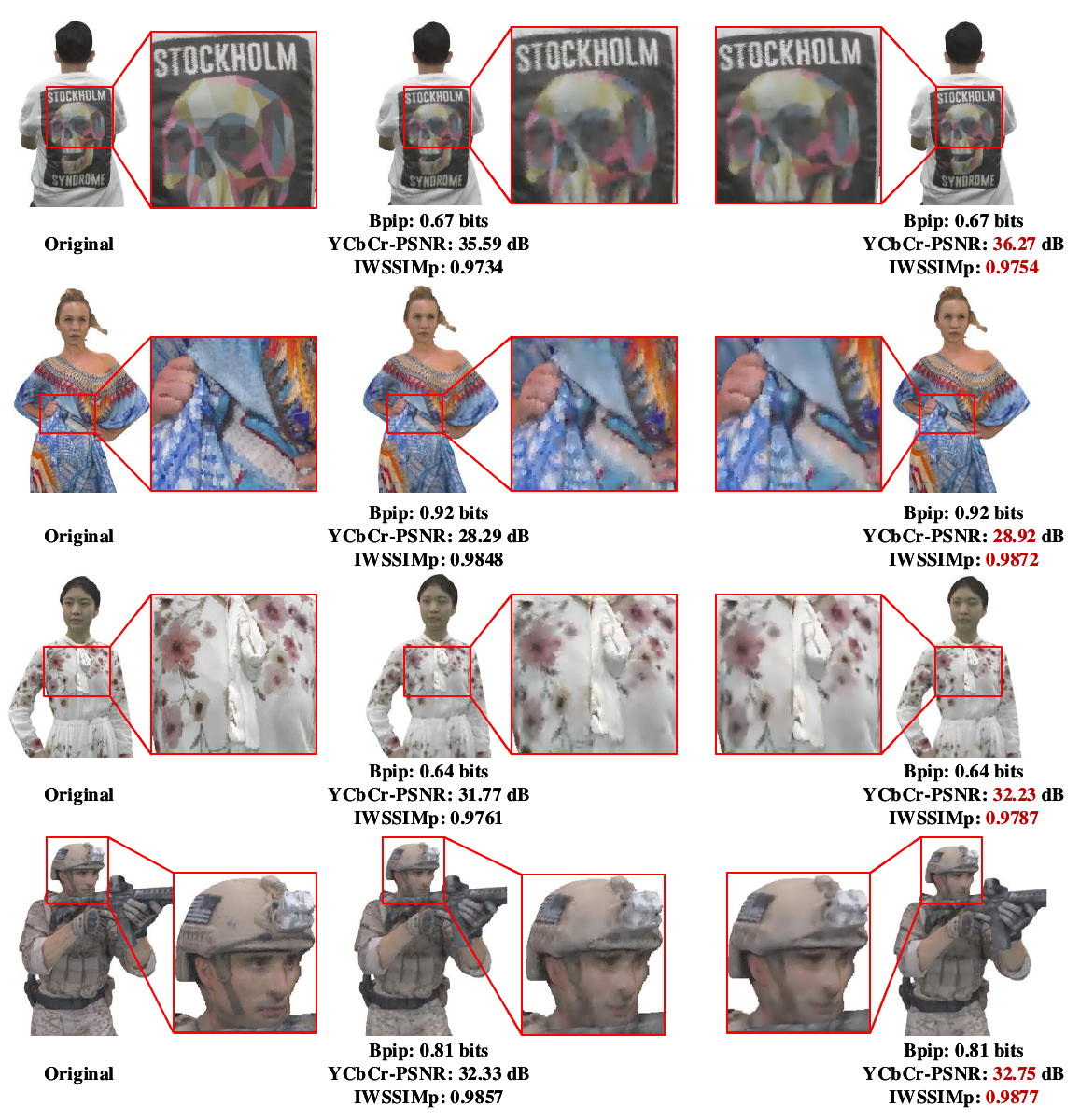}
\caption{Subjective quality comparison for the point clouds (from top to bottom) \textit{basketball\_player}, \textit{longdress}, \textit{model}, and \textit{soldier}. From left to right: original point clouds, point clouds compressed using G-PCC with the PredLift configuration, and point clouds enhanced by PCE-GAN.}
\label{FIG12}
\end{figure*}
\begin{figure*}
\centering
\includegraphics[width=4.6in]{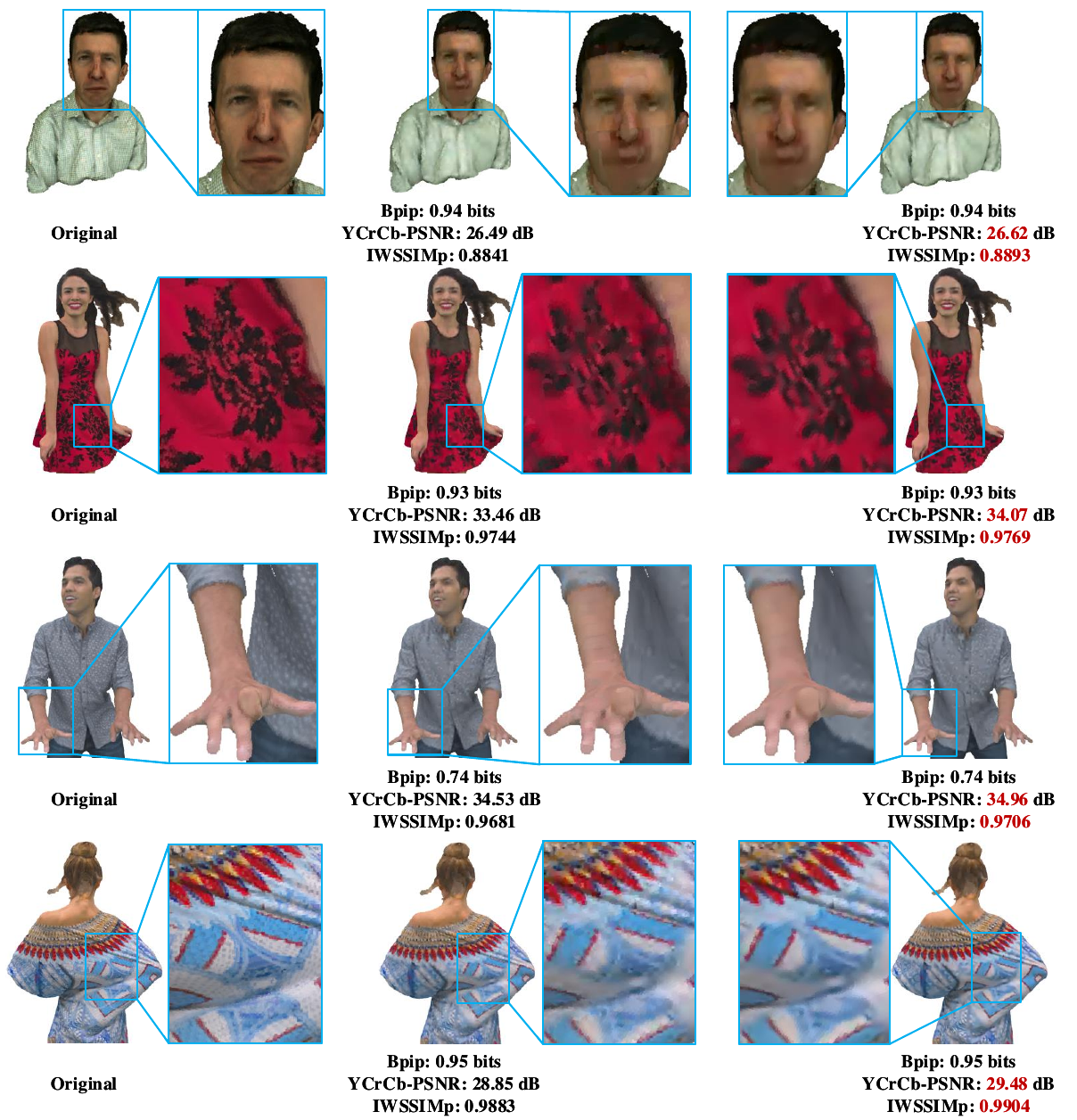}
\caption{Subjective quality comparison for the point clouds (from top to bottom) \textit{Andrew}, \textit{redandblack}, \textit{loot}, and \textit{longdress}. From left to right: original point clouds, point clouds compressed using G-PCC with the RAHT configuration, and point clouds enhanced by PCE-GAN.}
\label{FIG13}
\end{figure*}

Additionally, we visually compared the reconstructed point clouds with and without the proposed PCE-GAN as shown in Fig. 12 and Fig. 13. The point clouds processed by PCE-GAN showed a significant improvement in texture clarity and color transitions, revealing finer details and more natural color gradients. This enhanced not only the visual appearance of local regions but also effectively improved the overall visual experience.

\subsection{Robustness Analysis}
To further demonstrate the effectiveness of PCE-GAN, we applied the trained PCE-GAN models directly to a new dataset without retraining. We tested its performance on the broad quality assessment of static point clouds in a compression scenario (BASICS) \cite{ref51} dataset. The BASICS dataset consists of 75 point clouds, categorized into "\textit{Humans \& Animals}", "\textit{Inanimate Objects}", and "\textit{Buildings \& Landscapes}," with 25 point clouds in each category, encompassing a wide range of subjects such as animals, humans, everyday items, vehicles, architectural structures, and natural landscapes. All 75 point clouds were compressed using the G-PCC reference software TMC13v26 with the PredLift configuration at six bitrates, resulting in \(75\times6=450\) reconstructed point clouds. \begin{table}
\centering
\caption{RATE-\(\Delta\)PSNR AND PERCEPTUAL QUALITY COMPARISON WITH TMC13V26 LT ON BASICS}
\label{tab:my_label5}  
\resizebox{8.4cm}{!}{
\begin{tabular}{lccccc}
\toprule
\multirow{2}{*}{\textbf{BASICS}} & \multicolumn{4}{c}{\(\Delta\textbf{PSNR}\:\textbf{(dB)}\)} & \multirow{2}{*}{\(\Delta\textbf{IWSSIM}_{\textbf{P}}\:(\textbf{10}^{\textbf{-3}})\:\uparrow \)}\\ 
 & \textbf{Luma} & \textbf{Cb} & \textbf{Cr} & \textbf{YCbCr}  \\ \hline
        R01 & 0.357 & 0.148 & 0.116 & 0.329 & 3.4441 \\ 
        R02 & 0.479 & 0.218 & 0.224 & 0.447 & 2.7256 \\ 
        R03 & 0.601 & 0.299 & 0.303 & 0.564 & 1.6504 \\ 
        R04 & 0.618 & 0.453 & 0.447 & 0.597 & 0.9641 \\ 
        R05 & 0.540 & 0.545 & 0.578 & 0.543 & 0.4171 \\ 
        R06 & 0.382 & 0.657 & 0.711 & 0.419 & 0.0503 \\ \hline
        \textbf{Average} & \textbf{0.496} & \textbf{0.387} & \textbf{0.397} & \textbf{0.483} & \textbf{1.5419} \\ \bottomrule
\end{tabular}}
\end{table}
\begin{table}
\centering
\caption{ENHANCEMENT RESULTS FOR COMPLEX TEXTURE POINT CLOUDS IN BASICS DATASET}
\label{tab:my_label6}  
\resizebox{8.4cm}{!}{
\begin{tabular}{lccccc}
\toprule
\multirow{2}{*}{\textbf{Sequence}} & \multicolumn{4}{c}{\textbf{BD-PSNR} \textbf{(dB)}} & \multirow{2}{*}{\(\textbf{BD-IWSSIM}_{\textbf{P}}\:(\textbf{10}^{\textbf{-3}})\:\uparrow \)}\\ 
 & \textbf{Luma} & \textbf{Cb} & \textbf{Cr} & \textbf{YCbCr}  \\ \hline
        p12 & 1.019 & 0.361 & 0.517 & 0.947 & 0.7556 \\ 
        p13 & 1.064 & 0.498 & 0.772 & 1.010 & 1.2950 \\ 
        p14 & 1.314 & 1.070 & 0.790 & 1.266 & 0.9939 \\ 
        p30 & 1.302 & 0.730 & 0.337 & 1.206 & 1.5382 \\ 
        p37 & 0.869 & 0.621 & 0.690 & 0.842 & 1.8877 \\ 
        p38 & 0.998 & 0.373 & 0.276 & 0.914 & 3.2962 \\ 
        p42 & 1.136 & 0.717 & 0.114 & 1.046 & 2.2706 \\ 
        p44 & 0.948 & 0.365 & 0.324 & 0.873 & 2.8238 \\ 
        p60 & 1.037 & 0.353 & 0.382 & 0.954 & 2.1082 \\ 
        p74 & 0.865 & 0.542 & 0.445 & 0.818 & 2.8684 \\ \hline
        \textbf{Average} & \textbf{1.055} & \textbf{0.563} & \textbf{0.464} & \textbf{0.987} & \textbf{1.9837} \\ \bottomrule
\end{tabular}}
\end{table}We tested all point clouds, and the average results for both PSNR and \(\text{IWSSIM}_{\text{P}}\) are presented in Table V. PCE-GAN achieved an average improvement of 0.50 dB, 0.39 dB, and 0.40 dB in PSNR for the Y, Cb, and Cr, respectively. PCE-GAN showed a more significant improvement in PSNR at medium bitrates, while achieving better enhancement in \(\text{IWSSIM}_{\text{P}}\) at low bitrates. Moreover, to validate that PCE-GAN is better suited for processing point clouds with complex textures, we present in Table VI the results from several point clouds with intricate textures from the BASICS dataset. PCE-GAN achieved an average PSNR improvement of 1.06 dB, 0.56 dB, and 0.46 dB for the Y, Cb, and Cr components, respectively.
\begin{table}
\centering
\caption{TIME COMPLEXITY (TESTED ON NVIDIA RTX4090 GPU)}
\label{tab:my_label7}  
\resizebox{6cm}{!}{
\begin{tabular}{lccc}
\toprule
        \textbf{Sequence} & \textbf{Pat\_gen (s)} & \textbf{Process (s)} & \textbf{Pat\_fus (s)} \\ \hline
        basketball & 0.2243 & 197.7188 & 0.0013 \\ 
        dancer & 0.1692 & 173.9990 & 0.0010 \\ 
        exercise & 0.2270 & 163.5528 & 0.0010 \\ 
        longdress & 0.0706 & 56.6030 & 0.0011 \\ 
        loot & 0.0544 & 56.2321 & 0.0011 \\ 
        model & 0.1859 & 177.6594 & 0.0010 \\ 
        queen & 0.0902 & 65.5587 & 0.0009 \\ 
        redandblack & 0.0751 & 49.5852 & 0.0009 \\ 
        soldier & 0.0667 & 71.8271 & 0.0009 \\ 
        Andrew & 0.0991 & 86.1976 & 0.0013 \\ 
        David & 0.1229 & 85.9444 & 0.0008 \\ 
        Phil & 0.0907 & 96.7700 & 0.0008 \\ 
        Ricardo & 0.0592 & 61.4775 & 0.0009 \\ 
        Sarah & 0.1592 & 71.3309 & 0.0008 \\ \hline
        \textbf{Average} & \textbf{0.1210} & \textbf{101.0326} & \textbf{0.0010} \\ \bottomrule
\end{tabular}}
\end{table}

\subsection{Time Complexity Analysis}
Table VII shows the patch generation time (Pat\_gen), patch fusion time (Pat\_fus), and PCE-GAN processing time for all point clouds in the test dataset. To speed up patch generation and fusion, we used Boolean indexing assignment \cite{ref52}. This approach exploits vectorized operations to avoid explicit loops, significantly enhancing execution speed. 

\begin{table*}[ht]
\centering
\caption{QUANTITATIVE GAINS COMPARISON BETWEEN GQE-NET \cite{ref20} AND PCE-GAN}
\label{tab:my_label8}  
\resizebox{15cm}{!}{
\begin{tabular}{lcccccc|cccccc}
\toprule
\multirow{3}{*}{\textbf{Sequence}} & \multicolumn{6}{c}{\textbf{GQE-Net}} & \multicolumn{6}{c}{\textbf{PCE-GAN}}\\
& \multicolumn{4}{c}{\textbf{BD-PSNR} \textbf{(dB)}} & {\textbf{BD-rate}} & {\(\textbf{BD-IWSSIM}_{\textbf{P}}\)}& \multicolumn{4}{c}{\textbf{BD-PSNR} \textbf{(dB)}} & {\textbf{BD-rate}} & {\(\textbf{BD-IWSSIM}_{\textbf{P}}\)}\\ 
 & \textbf{Luma} & \textbf{Cb} & \textbf{Cr} & \textbf{YCbCr} & \textbf{(\%)}& (\(\textbf{10}^{\textbf{-3}}\))\(\uparrow\)&\textbf{Luma} & \textbf{Cb} & \textbf{Cr} & \textbf{YCbCr}& \textbf{(\%)}&(\(\textbf{10}^{\textbf{-3}}\))\(\uparrow\) \\ \hline
        basketball & 0.346 & 0.149 & 0.284 & 0.33 & -11.5 & 0.4953 & 0.460 & 0.190 & 0.423 & 0.441 & -15.0 & 0.7562 \\ 
        dancer & 0.385 & 0.140 & 0.297 & 0.364 & -11.9 & 0.6488 & 0.505 & 0.204 & 0.457 & 0.483 & -15.3 & 0.9186 \\ 
        exercise & 0.211 & 0.109 & 0.225 & 0.206 & -9.5 & 0.4431 & 0.280 & 0.139 & 0.323 & 0.274 & -12.6 & 0.6649 \\ 
        longdress & 0.319 & 0.228 & 0.332 & 0.314 & -10.0 & 1.2650 & 0.726 & 0.381 & 0.610 & 0.697 & -15.4 & 1.7669 \\ 
        loot & 0.319 & 0.229 & 0.276 & 0.311 & -9.6 & 0.6419 & 0.474 & 0.284 & 0.494 & 0.463 & -13.8 & 0.8753 \\ 
        model & 0.356 & 0.153 & 0.324 & 0.342 & -11.3 & 0.8257 & 0.554 & 0.197 & 0.509 & 0.529 & -14.7 & 1.1230 \\ 
        queen & 0.198 & -0.211 & -0.662 & 0.119 & -4.7 & 0.3859 & 0.363 & 0.578 & 0.653 & 0.394 & -10.8 & 0.6716 \\ 
        redandblack & 0.329 & 0.210 & 0.584 & 0.338 & -11.1 & 0.5843 & 0.562 & 0.225 & 0.708 & 0.550 & -15.1 & 0.7793 \\ 
        soldier & 0.416 & 0.211 & 0.263 & 0.394 & -11.5 & 1.0073 & 0.732 & 0.154 & 0.385 & 0.674 & -16.6 & 1.2959 \\ 
        Andrew & 0.516 & 0.160 & 0.014 & 0.462 & -15.9 & 0.7559 & 0.720 & 0.150 & 0.170 & 0.650 & -17.9 & 0.8901 \\ 
        David & 0.187 & 0.142 & 0.202 & 0.185 & -8.3 & 0.0824 & 0.167 & 0.127 & 0.303 & 0.173 & -6.8 & 0.2008 \\ 
        Phil & 0.519 & 0.122 & 0.045 & 0.464 & -15.4 & 0.8541 & 0.768 & 0.191 & 0.261 & 0.700 & -18.8 & 1.0024 \\ 
        Ricardo & 0.369 & 0.217 & 0.171 & 0.347 & -11.6 & 0.1634 & 0.450 & 0.224 & 0.319 & 0.427 & -14.0 & 0.2179 \\ 
        Sarah & 0.422 & 0.189 & 0.170 & 0.391 & -14.2 & 0.2606 & 0.567 & 0.193 & 0.271 & 0.525 & -15.5 & 0.3652 \\ \hline
        Average & 0.349 & 0.146 & 0.180 & 0.326 & -11.2 & 0.6010 & \textbf{0.523} & \textbf{0.231} & \textbf{0.420} & \textbf{0.499} & \textbf{-14.5} & \textbf{0.8234} \\ \bottomrule
\end{tabular}}
\end{table*}
\begin{table*}[ht]
\centering
\caption{ABLATION STUDY OF DGC, MGSA, FR, GFP AND GSCE MODULES AND NUMBERS OF NEIGHBOUR PATCHES IN TERMS OF \(\Delta\)PSNR}
\label{tab:my_label9}  
\resizebox{15cm}{!}{
\begin{tabular}{lcccccccc}
\toprule
        \textbf{Sequence} & \textbf{w/o DGC} &\textbf{ w/o MGSA} & \textbf{w/o FR} & \textbf{w/o GFP} & \textbf{w/o GSCE} & \({\bm{num}}_{\bm{nei}} = 4\) & \({\bm{num}}_{\bm{nei}} = 8\) & \textbf{PCE-GAN} \\ \hline
        basketball & 0.489 & 0.530 & 0.527 & 0.520 & 0.532 & 0.497 & 0.527 & 0.557 \\ 
        dancer & 0.545 & 0.588 & 0.597 & 0.577 & 0.591 & 0.554 & 0.583 & 0.617 \\ 
        exercise & 0.310 & 0.320 & 0.306 & 0.312 & 0.313 & 0.309 & 0.313 & 0.338 \\ 
        longdress & 1.126 & 1.157 & 1.003 & 1.037 & 1.144 & 1.172 & 1.175 & 1.234 \\ 
        loot & 1.239 & 1.249 & 1.162 & 1.209 & 1.222 & 1.252 & 1.241 & 1.360 \\ 
        model & 0.613 & 0.661 & 0.629 & 0.651 & 0.661 & 0.621 & 0.662 & 0.661 \\ 
        queen & 0.607 & 0.693 & 0.633 & 0.702 & 0.684 & 0.691 & 0.716 & 0.729 \\ 
        redandblack & 1.007 & 1.048 & 0.911 & 1.027 & 1.016 & 1.043 & 1.086 & 1.105 \\ 
        soldier & 1.531 & 1.543 & 1.448 & 1.582 & 1.548 & 1.553 & 1.581 & 1.664 \\ 
        Andrew & 1.504 & 1.469 & 1.215 & 1.437 & 1.469 & 1.483 & 1.475 & 1.611 \\ 
        David & 0.534 & 0.579 & 0.578 & 0.582 & 0.581 & 0.563 & 0.614 & 0.695 \\ 
        Phil & 1.264 & 1.317 & 1.078 & 1.328 & 1.342 & 1.308 & 1.343 & 1.436 \\ 
        Ricardo & 0.683 & 0.754 & 0.671 & 0.717 & 0.754 & 0.722 & 0.768 & 0.861 \\ 
        Sarah & 0.830 & 0.936 & 0.811 & 0.905 & 0.970 & 0.892 & 0.967 & 1.036 \\ \hline
        \textbf{Average} & 0.877 & 0.918 & 0.827 & 0.899 & 0.916 & 0.904 & 0.932 & \textbf{0.993} \\ \bottomrule
\end{tabular}}
\end{table*}
\subsection{Comparison with the State-of-the-Art}
We compared PCE-GAN with GQE-Net \cite{ref20}, the current state-of-the-art method. GQE-Net trains three separate models for the Y, Cb, and Cr components, with bitrates ranging from R01 to R06. Notably, GQE-Net does not require training a separate model for each bitrate. To ensure a fair comparison, we retrained our network following the training strategy of GQE-Net. The training and test point clouds were compressed using G-PCC TMC13v26 with PredLift according to the configuration adopted in \cite{ref20}. The PSNR and \(\text{IWSSIM}_{\text{P}}\) gains between GQE-Net and PCE-GAN are compared in Table VIII. PCE-GAN significantly outperformed GQE-Net in both PSNR and \(\text{IWSSIM}_{\text{P}}\) gains.

\subsection{Ablation Study}
To verify the effectiveness of proposed modules in PCE-GAN, we compared the performance of PCE-GAN with the following configurations:

\noindent (i) PCE-GAN \textbf{w/o DGC}: DGC was removed from PCE-GAN.

\noindent (ii) PCE-GAN \textbf{w/o MGSA}: we used an equivalent amount of MLPs to replace MGSA.

\noindent (iii) PCE-GAN \textbf{w/o FR}: FR was removed from PCE-GAN.

\noindent (iv) PCE-GAN \textbf{w/o GFP}: GFP was removed from PCE-GAN.

\noindent (v) PCE-GAN \textbf{w/o GSCE}: GSCE was removed from PCE-GAN.

\noindent (vi) \({\bm{num}}_{\bm{nei}} = 4\), i.e., four neighboring patches of the current patch were used to generate a grouped patch.

\noindent (vii) \({\bm{num}}_{\bm{nei}} = 8\), i.e., eight neighboring patches of the current patch were used to generate a grouped patch.

Table IX shows the comparison results. All the modules (\textbf{DGC}, \textbf{MGSA},\textbf{FR}, \textbf{GFP}, and \textbf{GSCE}) contributed to the overall performance of PCE-GAN. Notably, PCE-GAN used the geometry information of neighboring patches to capture global correlations. When the number of patches was too small (e.g., \({num}_{nei} = 4\)), the available geometry information was insufficient, leading to poor global correlation and reduced network performance. Conversely, when the number of patches was too large (e.g., \({num}_{nei} = 8\)), the KNN search resulted in an excessive number of duplicate points, which degraded performance. The selected patch number, \({num}_{nei} = 6\), allowed the neighboring patches to effectively surround the current patch while avoiding information redundancy.

\section{Conclusion}
We proposed a point cloud attribute quality enhancement algorithm based on optimal transport theory. Our method ensures the fidelity of the enhanced point cloud while achieving high perceptual quality by constraining the enhanced attributes to follow the same distribution as the original ones. Specifically, we formulated the quality enhancement problem as an optimal transport problem and developed a relaxation strategy for its solution by introducing a generative adversarial network (PCE-GAN). The generator primarily consists of a local feature extraction unit, a global spatial correlation unit, and a feature squeeze unit. The discriminator incorporates a WGAN-GP structure to compute the Wasserstein-1 distance between the enhanced point cloud and the original point cloud. To the best of our knowledge, this is the first exploration of point cloud attribute quality enhancement that explicitly considers perceptual quality. Experimental results showed that PCE-GAN significantly improved the quality of compressed point cloud attributes. In particular, the results showed remarkable superiority of PCE-GAN in processing point clouds with complex textures, better aligning with the needs of real-world applications. In the future, we aim to reduce computational complexity and further enhance quality.

\newpage

\vfill

\end{document}